\documentclass{article}
 \usepackage{color}
\usepackage{amsfonts}
\usepackage{amsmath}
\usepackage{amssymb}
\usepackage{graphicx}
\usepackage{comment}

\newtheorem{theorem}{Theorem}

\newcommand{\ba}{\begin{eqnarray}}
\newcommand{\ea}{\end{eqnarray}}
\newcommand{\nn}{\nonumber}

\newcommand{\cF}{{\mathcal{F}}}

\newcommand{\be}{\begin{equation}}
\newcommand{\ee}{\end{equation}}

\newcommand{\cS}{\mathcal{S}}
\newcommand{\cD}{\mathcal{D}}

\def\a{\alpha}
\def\b{\beta}

\def\d{\delta}
\def\e{\epsilon}

\def\k{\kappa}
\def\l{\lambda}

\def\x{\xi}

\def\D{\Delta}

\def\L{\Lambda}

\allowdisplaybreaks[1]

\begin{document}
%
%
%
%
%
\begin{titlepage}

\begin{flushright}
UT-14-23
\end{flushright}

\vskip 12mm

\begin{center}
{\Large Construction of Gaiotto states with fundamental
multiplets}\\
{\Large through Degenerate DAHA}
\vskip 2cm
{\Large Yutaka Matsuo$^*$, Chaiho Rim$^\dagger$  and Hong Zhang$^\dagger$}
\vskip 2cm
$^*$ {\it Department of Physics, The University of Tokyo, Bunkyo-ku, 
Tokyo, Japan}\\

$^\dagger$ {\it Department of Physics and Center for Quantum Spacetime (CQUeST)}\\
{\it Sogang University, Seoul 121-742, Korea}
\end{center}
\vfill
\begin{abstract}
We construct Gaiotto states with fundamental multiplets in $SU(N)$ gauge theories, in terms of the orthonormal basis of spherical degenerate double affine Hecke algebra (SH in short), the representations of which are equivalent to those of $W_n$ algebra with additional $U(1)$ current. The generalized Whittaker conditions are demonstrated under the action of SH, and further rewritten in terms of $W_n$ algebra. Our approach not only consists with the existing literature but also holds for general $SU(N)$ case.
\end{abstract}
\vfill
\end{titlepage}

\setcounter{footnote}{0}

\section{Introduction}
The instanton Nekrasov partition function\cite{r:Nekrasov, 0306238, r:NY} for 4 dimensional  ${\cal N} =2 $
supersymmetric $SU(2)$ quiver gauge theory has the remarkable correspondence with 
2 dimensional Liouville conformal field theories,
so called AGT conjecture \cite{AGT}. 
And the correspondence was generalized into $SU(N)$ quiver gauge theories  in \cite{Wyllard:2009hg, Mironov:2009by} . 

There are various proofs for the AGT conjecture \cite{DAHA-AGT, Morozov:2013rma,
 Itoyama:2013mca, Alba:2010qc, Fateev:2011hq}. 
In most cases, the Virasoro and W algebras play the essential role. 
In contrast, spherical degenerate
double affine Hecke algebra (spherical DDAHA or SH)  
\cite{r:DAHA, Maulik:2012wi, Smirnov:2013hh,r:SV, Kanno:2013aha}
turns out to be another useful tool to prove the AGT conjecture. 
DDAHA is generated by $2N$ operators,
$z_i$ and $\cD_i$ ($i=1,\cdots,N$) 
where 
\begin{eqnarray}
\cD_i=z_i \nabla_i+\sum_{j<i} \sigma_{ij},\quad
\nabla_i=\frac{\partial}{\partial z_i}
+\beta\sum_{j(\neq i)}\frac{1}{z_i-z_j}(1-\sigma_{ij})\,.
\end{eqnarray} 
and permutation operators. Here $\nabla_i$ is the Dunkl operator which 
 plays a fundamental role in Calogero-Sutherland system
 and $\sigma_{ij}$ is the transposition of variables, 
$ z_i \sigma_{ij} = \sigma_{ij}z_j $. 
The operators $z_i$ and $\cD_i$ satisfies the following commutation relations,
\ba
&&[z_i, z_j]=0,\qquad [\cD_i, \cD_j]=0\,,\\
&& [\cD_i, z_j]=\left\{
\begin{array}{ll}
-\beta z_i \sigma_{ij} \quad & i<j\\
z_i+\beta \Big (\sum_{k<i} z_k \sigma_{ik}+\sum_{k>i} z_i\sigma_{ik} 
\Big)  \quad & i=j\\
-\beta z_j \sigma_{ij}\quad & i>j\,.
\end{array}
\right.
\ea 
DDAHA is the algebra freely 
generated by  $z_i, D_i$ and $\sigma\in S_N$. 
Spherical DDAHA (SH) is obtained by the restriction to the symmetric part.
For the special value of $\beta=1$, SH reduces to
$\mathcal{W}_{1+\infty}$ algebra which is described by free fermions.

Recently it was found that some representations of SH 
are equivalent to those of $W_n$ algebra with additional $U(1)$ current
\cite{r:SV}. 
It is known that SH has a natural action on the equivariant cohomology class
of the instanton moduli space while $W_n$ algebra describes the 
symmetry of Toda field theory.  
This correspondence
was used to prove the AGT conjecture.
For example,  in \cite{r:SV} such mechanism was applied to the pure 
$SU(N)$ super Yang-Mills theory, and 
the representative of the cohomology class is mapped to
the orthogonal basis in the Hilbert space of $W_n$ algebra.
In this way, the Gaiotto state \cite{r:Gaiotto} is constructed
to arbitrary order through the conditions on the action of the generators of SH.
Later in  \cite{Kanno:2013aha}, 
such correspondence was
applied to quiver type gauge theories.
The action of SH on the basis appears as the
recursion relation for the Nekrasov partition function, which is then interpreted as the Ward identities
associated with the $W_n$-algebra.

Here we apply the similar trick to 
 construct explicit
Gaiotto states with fundamental  multiplets in $SU(N)$ gauge theories. 
The computation is in parallel with those in \cite{r:SV}. 
Note that the Gaiotto state appears as an irregular module of
Virasoro and $W_n$ algebra. There were already a few 
attempts to construct the irregular states algebraically  in 
\cite{r:Gaiotto, irregular, Kanno:2012}. Our construction is not limited to $SU(3)$ but is 
extended to   $SU(N)$ with $N_f < N$. 

It is also noted that the Gaiotto state construction was proposed but in a 
different manner, which uses the coherent state approach 
in \cite{GT2012, KMST2013}. Some of  irregular state was
constructed explicitly using random matrix formalism 
in connection with $SU(2)$ quiver gauge theories \cite{RNC}.
Thus, our construction will be instructive and complementary 
to understand the Gaiotto state in different approaches. 

This paper is organizes as follows. 
 In section \ref{sec:Gaiotto} we define the Gaiotto states
with fundamental multiplets in terms of the orthonormal basis of SH.
In section \ref{sec:intro-SH}, we briefly review the algebra SH
and the relation with $W_n$ algebra.
In section \ref{sec:SH-W}, we give the explicit correspondence
between SH and $W_n$ generators through the use of free boson
fields.  In section \ref{sec:Whi} we show that the states satisfy generalized 
Whittaker condition in terms of SH.  Finally in section \ref{sec:Whi-W}
we rewrite the conditions in terms of the generators of $W_n$
algebra and confirm the consistency with the existing literature
\cite{irregular,  Kanno:2012, GT2012}.

In the appendix, we derive the Ward identities for the Virasoro operator
$L_{\pm 2}$. Though this is not directly relevant to the main claim of this paper,
we include it since the analysis is technically very close
and also it completes the analysis of \cite{Kanno:2013aha}.

\section{Construction of Gaiotto States}
\label{sec:Gaiotto}
%
For the pure super Yang-Mills theory where the fundamental multiplet is absent, $N_f=0$, 
the instanton part of the partition function has the form,
\be
Z(\vec a) =\sum_{\vec Y}\L^{4|\vec Y|}Z_{\text {vect}}(\vec a, \vec Y) ,
\ee
with
\be
Z_{\text {vect}}(\vec a, \vec Y):=f(\vec a, \vec Y):=
 \prod_{p,q}\frac{1}{
 g_{Y_p Y_q}(a_p-a_q)}
\ee
where $\L$ is the dynamical scale, 
$\vec a\in \mathbf{C}^n$  is the VEV for an adjoint scalar field in the vector multiplet
and $\vec Y=(Y_1,\cdots, Y_N)$ is a set of Young tableaux 
characterizing fixed points of localization in the  instanton moduli space. And
\ba
g_{Y,W}(x)&=&\prod_{(i,j)\in Y}(x+\beta(Y^\prime_j-i+1)+W_i-j)
\prod_{(i,j)\in W}(-x+\beta(W^\prime_j-i)+Y_i-j+1)\,,
\ea
 where $Y_i$ is the $i$th column of $Y$, and
 $Y^\prime$ stands for the transposed Young tableaux.  $\beta$ is related to 
$\Omega$-deformation parameters by $\beta=-\e_1/\e_2$.

According to AGT conjecture, we may put  the partition function  
as the inner product of two  Gaiotto states 
$Z(\vec a) =\langle \tilde G |G\rangle $.
It is a nontrivial issue to realize $ |G\rangle $ in the Hilbert space of W-algebra.  On the other hand, in SH, we know the orthonormal basis and the action of generators which will be reviewed in the next section. The Gaiotto state takes the form,
\ba
|G\rangle&=&\sum_{\vec Y}\L^{2|\vec Y|}(Z_{\text {vect}}(\vec a,\vec Y)
)^{1/2}|\vec a,\vec Y\rangle\,.
\ea
Here $ |\vec a,\vec Y\rangle$ is introduced in \cite{Kanno:2013aha} as an basis of a Hilbert space $\mathcal{H}_{\vec a}$. The dual basis $\langle \vec a, \vec Y|$ is defined such that 
$ 
\langle \vec a, \vec Y|\vec a, \vec W\rangle =\delta_{\vec Y, \vec W}\,.
$ It is trivial to confirm that it has the desired inner product
due to the orthonormal property of the basis. 
However, it is nontrivial to confirm that it satisfies the condition
for generalized Whittaker condition as given in \cite{r:SV}.

One may proceed likewise for $N_f=2$. 
The partition function has extra contributions from the fundamental multiplets with masses
 $m_i$,
\be
Z^{N_f=2}(\vec a, m_1, m_2,\L) =\sum_{\vec Y}\L^{2|\vec Y|}
Z_{\text {vect}}(\vec a, \vec Y)Z_{\text {fund}}(\vec a, \vec Y,m_1 )Z_{\text {fund}}(\vec a, \vec Y,m_2 )
\ee
where
\be
Z_{\text {fund}}(\vec a, \vec Y, m)=
\prod_{p=1}^N \prod_{(i,j)\in Y_p}(a_p+\b i -j -m)
\ee
Noting that 
\be
Z^{N_f=2}(\vec a, m_1, m_2,\L) =\langle G, m_2 |G, m_1\rangle
\ee
one may have the Gaiotto state with one additional parameter $m$
\ba
|G , m\rangle&=&\sum_{\vec Y}\L^{|\vec Y|}(Z_{\text {vect}}(\vec a,\vec Y))^{1/2}Z_{\text {fund}}(\vec a, \vec Y, m) \,|\vec a,\vec Y\rangle\,.
\ea
In this way,
it is straightforward to generalize it to additional $k<N$ parameters $m_1, m_2, \cdots, m_k$,
namely,
\ba
|G , m_1,\cdots,m_{k}\rangle&=&\sum_{\vec Y}\L^{|\vec Y|}(Z_{\text {vect}}(\vec a,\vec Y))^{1/2}\prod_{A=1}^{k} (Z_{\text {fund}}(\vec a, \vec Y, m_A)) |\vec a,\vec Y\rangle\,.
\ea
One may easily confirm that the inner product of two Gaiotto states with $k$ parameters 
will give the instanton partition function with $N_f=2k$. The nontrivial part is to confirm the Whittaker vector conditions. The case for $N_f=0$ was given by \cite{r:SV}.
The proof for additional fundamental multiplets is new.
Our task is to find the generalized Whittaker conditions using SH generators 
and rewrite them in terms of $W_n$ generators.
\section{Brief introduction of SH}
\label{sec:intro-SH}
The generators of Spherical DDAHA (SH) are obtained 
by symmetrizing those of DDAHA by $\cS=\frac{1}{N!} \sum_{\sigma\in S_N} \sigma$,
$
\cS \mathcal{O} \cS$
where $\mathcal{O}\in \mbox{DDAHA}$.
Such generators act naturally on the ring of 
symmetric functions of $z_i$.
The independent generators of SH is given by
$D_{nm}\sim \cS \sum_{i=1}^N (z_i)^n (\cD_i)^m\cS$
($n\in \mathbf{Z}$, $m\in \mathbf{Z}_{\geq 0}$)
in $N\rightarrow \infty$ limit.
The definition of $D_{nm}$
is only sketchy here and will be more carefully defined later.
For a special value for $\beta=1$, SH reduces to
$\mathcal{W}_{1+\infty}$ algebra which is described by free fermions.

In large $N$ limit, one may introduce free boson description of SH
in terms of power sum polynomial $p_n=\sum_{i=1}^\infty (z_i)^n$.
We identify,
\ba
p_n:= \a_{-n},\quad
n\frac{\partial}{\partial p_n}:= \a_{n},\quad n\in \mbox{Z}_{\geq 0}\,
\ea
which satisfies the standard commutation relation $[\a_n,\a_m]=n \delta_{n+m,0}$.
The space of symmetric functions is described by the Fock space
$\cF$ of the free boson.

The Hilbert of $W_n$-algebra shows up when we take coproduct
of $n$ representations of $\cF$ and make some restriction on
the representation (taking the `symmetric part' which is referred
as $[1^n]$ representation in \cite{r:SV}). After taking
such coproduct it has nontrivial central charges given below.
To distinguish the algebra with central extensions from others, we will denote 
the algebra SH$^c$.
It has generators $D_{r,s}$  with $r\in \mathbf{Z}$ and $s\in \mathbf{Z}_{\geq 0}$. 
The commutation relations for degree $\pm 1, 0$ generators are defined by,
\ba
\left[D_{0,l} , D_{1,k} \right] & =& D_{1,l+k-1}, \;\;\; l \geq 1 \,,\label{SH1}\\
\left[D_{0,l},D_{-1,k}\right]&=&-D_{-1,l+k-1}, \;\;\; l \geq 1 \,,\\
\left[D_{-1,k},D_{1,l}\right]&=&E_{k+l} \;\;\; l,k \geq 0\label{eDDE}\,,\\
\left[D_{0,l} , D_{0,k} \right] & =& 0 \,,\,\, k,l\geq 0\,,\label{SH4}
\ea
where 
$E_k$ is a nonlinear combination of $D_{0,k}$ determined in the form of a generating function,
\ba
1+(1-\beta)\sum_{l\geq 0}E_l s^{l+1}= \exp(\sum_{l\geq 0}(-1)^{l+1}c_l \pi_l(s))\exp(\sum_{l\geq 0}D_{0,l+1} \omega_l(s)) \,,\label{com0}
\ea
with 
\ba
&&\pi_l(s)=s^l G_l(1+(1-\beta)s) \,,\\
&&\omega_l(s)=\sum_{q=1,-\beta,\beta-1}s^l(G_l(1-qs)-G_l(1+qs)) \,,\\
&&G_0(s)=-\log(s), \;\;\; G_l(s)=(s^{-l}-1)/l \;\;\; l \geq 1\,.
\ea
The parameters $c_l$ ($l\geq 0$) are central charges.
Other generators are defined recursively by,
\ba
D_{l+1,0} = \frac1l \left[D_{1,1} , D_{l,0} \right] ,&\qquad&
D_{-l-1,0} = \frac1l \left[D_{-l,0},D_{-1,1}\right] \,, \\
D_{r,l} = \left[D_{0,l+1} , D_{r,0} \right]    \;\;\;  &\qquad&
D_{-r,l}= \left[D_{-r,0} , D_{0,l+1} \right]\,. 
\ea
for $l\geq 0, r>0$\,.

There is an
explicit form of the action on the orthonormal basis $|\vec a, \vec Y\rangle$,
\ba
D_{-1,l}|\vec a,\vec Y\rangle
&=&(-1)^{l} \sum_{q=1}^{N} \sum_{t=1}^{{f_q}}(a_q+B_t(Y_q))^l  \Lambda^{(t,-)}_q(\vec Y)|\vec a,\vec Y^{(t,-),q}\rangle\label{DW1} \,,\\
D_{1,l}|\vec a,\vec Y\rangle&=&(-1)^{l}\sum_{q=1}^{N}\sum_{t=1}^{{f_q}+1}(a_q+A_t(Y_q))^l  
\Lambda^{(t,+)}_q(\vec Y)|\vec a,\vec Y^{(t,+),q}\rangle\label{DW2}\,,\\
D_{0,l+1}|\vec a,\vec Y\rangle&=&(-1)^l \sum_{q=1}^{N}
\sum_{\mu \in Y_q}(a_q+c(\mu))^l |\vec a,\vec Y\rangle\,. \label{eD0W}
\ea
where
$
c(\mu)=\beta i-j \mbox{ for }\mu=(i,j).   
$
\begin{figure}[bpt]
\begin{center}
\includegraphics[scale=0.6]{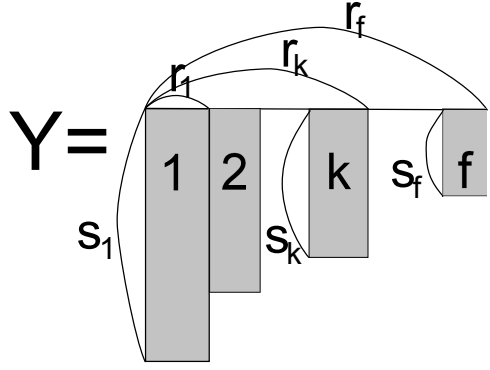}
\end{center}
\caption{Decomposition of Young diagram by rectangles}
\label{f:young}
\end{figure}
The factor $\Lambda^{(t,-)}_q(\vec a, \vec Y)$ is defined by
\ba
\Lambda^{(k,+)}_p(\vec a,\vec Y) &=& \left(
\prod_{q=1}^N \left(\prod_{\ell=1}^{f_q} \frac{
a_p-a_q+A_k(Y_p)-B_\ell(Y_q)+\xi
}{
a_p-a_q+A_k(Y_p)- B_\ell(Y_q)
}{\prod}_{\ell=1}^{\prime f_q +1}\frac{a_p-a_q+A_k(Y_p)- A_\ell(Y_q) -\xi}{
a_p-a_q+A_k(Y_p)-A_\ell(Y_q)
}
\right)\right)^{1/2}\,,\\
\Lambda^{(k,-)}_p(\vec a,\vec Y) &=& \left(
\prod_{q=1}^N \left(\prod_{\ell=1}^{ f_q+1} \frac{ a_p-a_q+B_k(Y_p)-A_\ell(Y_q)-\xi
}{a_p-a_q+B_k(Y_p)- A_\ell(Y_q)
}{\prod}_{\ell=1}^{\prime  f_q}\frac{a_p-a_q+B_k(Y_p)-B_\ell(Y_q)+\xi}{
 a_p-a_q+B_k(Y_p)-B_\ell(Y_q)
}
\right)\right)^{1/2}\,.
\ea
We decompose $Y$ into rectangles $Y=(r_1, \cdots, r_f; s_1,\cdots, s_f)$ 
(with $0<r_1<\cdots <r_f$, $s_1>\cdots>s_f>0$, see Figure \ref{f:young}
for the parametrization). 
We use $f_p$ (resp. $\bar f_p$) to represent the number of rectangles
of $Y_p$ (resp $W_p$).
The factors $A_k(Y_p)$, $B_\ell(Y_q)$ are\begin{eqnarray}
A_k(Y)& =&  \beta r_{k-1}-s_k-\xi,\quad (k=1,\cdots, f+1)\label{e:Ak}\,,\\
B_k(Y)&= &  \beta r_{k}-s_k,\quad (k=1,\cdots, f)\label{e:Bk}\,,
\end{eqnarray}
\begin{figure}[bpt]
\begin{center}
\includegraphics[scale=0.6]{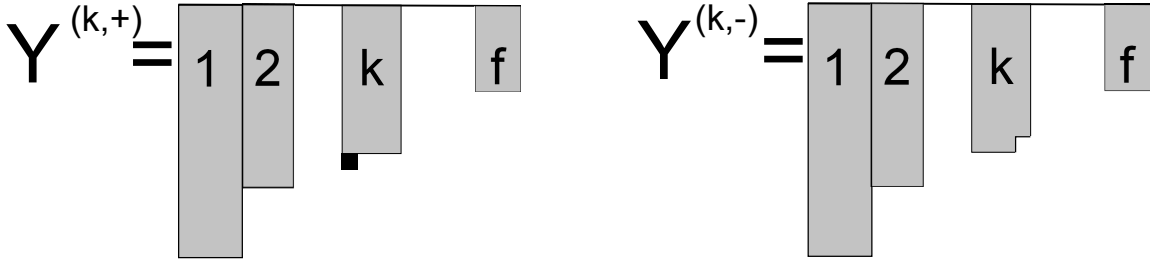}
\end{center}
\caption{Locations of boxes}
\label{f:Young+-}
\end{figure}
where $\xi:=1-\beta$.
$A_k(Y)$ (resp. $B_k(Y)$) represents the $k^\mathrm{th}$ location 
where a box may be added to (resp. deleted from) the Young diagram $Y$ 
composed with a map from location to $\mathbf{C}$. \\

We denote $Y^{(k,+)}$ (resp. $Y^{(k,-)}$) as the Young diagram obtained from
$Y$ by adding (resp. deleting) a box at $(r_{k-1}+1,s_k+1)$ (resp. $(r_k,s_k)$).
Similarly we use the notation $\vec Y^{(k\pm),p}=(Y_1,\cdots, Y_p^{(k,\pm)}, \cdots, Y_N)$
to represent the variation of one Young diagram in a set of Young tables $\vec Y$.
For more detail of the notation, we refer \cite{Kanno:2013aha}.

\section{The relation between SH$^c$ and $W$-algebra}
\label{sec:SH-W}
SH$^c$ and $W_n$-algebra look very different but the Hilbert space
of both algebras are identical for $[1^n]$ representation of SH$^c$.
The content of this section is a brief summary of \cite{r:SV}.

The generators of $W_n$-algebra are defined through the quantum 
Miura transformation,
\ba\label{e:Miura}
-: \prod_{j=1}^n (Q\partial_z +\vec h_j\cdot\partial\vec\varphi):
=\sum_{d=0}^n W^{(d)}(z)( Q\partial_z)^{n-d}\,.
\ea
where $\vec h_i =\vec e_i -\frac{1}{n} \sum_{i=1}^n \vec e_i$
and $\vec e_i$ is the $i$-th orthonormal basis of $\mathbf{R}^n$.
$\partial\vec\varphi=(\partial\varphi_1,\cdots,\partial\varphi_n)$
is $n$ free bosons with the standard OPE,
\ba
\partial\varphi_i(z) \partial\varphi_j(0)\sim \frac{\delta_{ij}}{z^2}\,,\quad
\partial \varphi_i(z) =\sum_{r\in \mathbf{Z}} \a_r^{(i)}z^{-r-1}\,.
\ea
We introduce $\mathcal{J}(z)=\sum_{i=1}^n \partial\varphi_i(z)$
to describe the $U(1)$ factor.

Expansion of (\ref{e:Miura}) gives,
\ba
W^{(0)}(z)&=&-1 ,\\
W^{(1)}(z)&=& 0,\\
W^{(2)}(z) &=& \frac12  (\partial\vec\varphi)^2
-\frac{1}{2n} :\mathcal{J}^2(z): +Q \vec\rho\cdot\partial^2\vec\varphi,
\label{WL2}
\ea
with $\vec\rho=(-\frac{n-1}{2},-\frac{n-3}{2},\cdots, \frac{n-1}{2})$.
$W^{(2)}$ is the standard form of Virasoro generators with the
central charge, $c=(n-1)(1-Q^2 n(n+1))$.  The higher generators
are in general complicated but the part with highest power of
$\partial\varphi$ is written in a relatively simple way,
\ba
W^{(d)}&=&-\sum_{j_1<\cdots <j_d} :(\vec h_{j_1} \cdot \partial \vec\varphi) \cdots 
(\vec h_{j_d} \cdot  \partial\vec \varphi):+
\mbox{ lower terms}\nonumber\\
&=&- \sum_{s=0}^d (-n)^{s-d} \left(
\begin{array}{c}
n-s \\ n-d
\end{array}
\right)\sum_{j_1<\cdots<j_s} : \mathcal{J}(z)^{d-s} \partial\varphi_{j_1}(z)
\cdots \partial\varphi_{j_s}(z)+\mbox{ lower terms}\,.
\ea

Meanwhile, SH$^c$ is given in free boson representation, obtained from
the expression for $D_{\pm 1,0}$ and $D_{0,2}$. 
For $[1^n]$ representation, they are
\ba
D_{\pm 1,0}&=& -\sum_{i=1}^n \a_{\mp 1}^{(i)}\,,\\
D_{0,2}&=& \sum_i^n\left\{ \frac{\sqrt{\beta}}{6}\sum_{r,s\in \mathbf{Z}} (:\a^{(i)}_r \a^{(i)}_s \a^{(i)}_{-r-s}:)
+\frac{\x}{2}\sum_{r>0} (r+1-2i) \a^{(i)}_{-r} \a^{(i)}_r
\right\}
+\x\sum_{i<j}^n  \sum_{r>0} r \a^{(i)}_{-r} \a^{(j)}_r \,.
\ea
While $D_{\pm 1,0}$ is diagonal with respect to the sum over $i$,
there exist off-diagonal term in $D_{0,2}$ which represents
the nontrivial twist in the coproduct. $D_{0,2}$ for $n=1$ case is identical to the Hamiltonian of Calogero-Sutherland \cite{Awata:1994xd}.

Generators of Heisenberg ($J_l$) and Virasoro algebras ($L_l$) 
are embedded in SH$^c$ as \cite{r:SV},
\ba
&& J_{l}=(-\sqrt{\beta})^{-l} D_{-l,0},\quad J_{-l}=(-\sqrt{\beta})^{-l} D_{l,0}, \quad J_0=E_1/\beta,
\label{defJ}\\
&& L_l=(-\sqrt{\beta})^{-l} D_{-l,1}/l +(1-l) c_0 \xi J_l/2\,,\quad\nn\\
&& L_{-l}= (-\sqrt{\beta})^{-l} D_{l,1}/l +(1-l)c_0 \xi J_{-l}/2\,,\nn\\
&& L_0=[L_1,L_{-1}]/2=D_{0,1} +\frac{1}{2\beta}\left(c_2+c_1(1-c_0) \xi +\frac{\xi^2}{6} c_0(c_0-1)(c_0-2)\right)\,,\label{defV}
\ea
where $
c_l=\sum_{p=1}^{N}(a_p -\xi)^l$ when act on $|\vec a, \vec Y\rangle$. The elements $D_{l,1}$ are obtained from the commutation relation,
$
D_{\pm r,1}= \pm [D_{0,2},D_{\pm r,0}]\,.
$ Here $J(z)=\frac{1}{\sqrt{\beta}}\sum_{i=1}^n \partial\varphi_i(z)$, and one may evaluate the Virasoro generator as,
\ba
L_n &=& \frac12 \sum_{i} \sum_m :\a^{(i)}_{n+m}  \a^{(i)}_{-m}:
+ Q \sum_{i} n\rho_i \a^{(i)}_n \,.
\ea
This agrees with the Virasoro generator in \eqref{WL2}
(with the contribution from $U(1)$ factor).
It implies that the Hilbert space of the $W_n$ algebra with $U(1)$ factor 
coincides with the $[1^n]$ representation of SH$^c$.

In the following, we derive the explicit form of the some generators of SH which are used in the next sections. The relation between higher generators can be similarly obtained using the commutators.  The procedure is simplified once
we compare the terms with highest generators.  For such purpose
it is more convenient to introduce a new 
set of elements $Y_{l,d}$ which 
are defined inductively starting from $Y_{\pm 1,d}=D_{\pm 1,d}$.
 For $l \geq  2$ and $d \geq 1$,
\begin{equation}\label{yrd}
Y_{l,d}=\begin{cases} [D_{1,1}, Y_{l-1,d}] & \text{if}\; l -1\neq d\\ 
[D_{1,0}, Y_{l-1,d+1}] & \text{if}\; l-1=d, \end{cases} \qquad
Y_{-l,d}=\begin{cases} [D_{-1,1}, Y_{1-l,d}] & \text{if}\; l-1 \neq d\\ 
[D_{-1,0}, Y_{1-l,d+1}] & \text{if}\; l-1=d,\end{cases}
\end{equation}

There exists a constant $c(l,d) \neq 0$ such that
\begin{equation}
Y_{l,d}\equiv
\label{Yboson}
c(l,d) \sum_{i=1}^r \,\,\sum_{ l_0+\ldots+l_d=-l} :  \a^{(i)}_{l_0} \cdots \a^{(i)}_{l_d}:
+\mbox{ lower order terms}.
\end{equation}
In particular,
\be
c(0,d)=\frac{\sqrt{\beta}^{d-1}}{d(d+1)},
\ee
and 
\begin{equation}
c(1,d)=-\sqrt{\beta}^d/(d+1),\qquad c(-1,d)=-\sqrt{\beta}^d/(d+1).
\end{equation}
The other coefficients are determined recursively.

Here we introduce a notation which is useful later.
Let $f(z_1, \ldots, z_n)=
\sum_{\underline{i}} a_{\underline{i}} z_1^{i_1} \cdots z_r^{i_n}$ 
is a symmetric polynomial with respect to $n$ variables
$z_1,\cdots, z_n$.  We will also denote the n-powers of bosonic fields with coefficients 
$a_i$ by 
\ba
:\!f (\underline{z}) \!:=\sum_{\underline{i}} 
a_{\underline{i}} :(\partial\varphi_1(z))^{i_1} \cdots (\partial\varphi_n(z))^{i_n}\!:
\ea
Furthermore we use a notation $\big(u(z)\big)_i=u_i$ when $u(z)$ 
with conformal dimension $d$ has the 
expansion $u(z)=\sum_i u_i z^{-i-d}$. With this preparation,
we use the power sum polynomial $p_l(z)=\sum_i (z_i)^l$
to represent the first few generators in a compact form,
\begin{equation}
D_{-1,d}\sim\frac{-\sqrt{\beta}^d}{d+1} \big( :\! p_{d+1}(\underline{z})\!:\big)_{1}, \qquad 
D_{0,d}\sim\frac{\sqrt{\beta}^{d-1}}{d(d+1)} \big( :\! p_{d+1}(\underline{z})\!:\big)_{0}.
\end{equation}
Here $\sim$ is used to imply that we neglect lower powers of $\partial\varphi$.
The next generator $D_{-2,d} $has the form:
\ba
\label{Dr2}
D_{-2,d} \sim\frac{2\sqrt{\beta}^{d+1}}{d+1} \big( :\! p_{d+1}(\underline{z})\!:\big)_{2}
\ea
which will be used in the next sections.
Here is an explicit proof of (\ref{Dr2}). 
We start with
\begin{equation}
D_{-1,d}\sim
c(-1,d) \sum_{i=1}^r \,\,\sum_{ l_0+\ldots+l_d=1} :  \a^{(i)}_{l_0} \cdots \a^{(i)}_{l_d}:.
\end{equation}
By $[\a_n,\a_m]=n\d_{n+m}$, we obtain
\begin{equation}
[D_{-1,1},D_{-1,d}]= Y_{-2,d}\sim
c(-1,1)\,c(-1,d) \times 2(d-1)\sum_{i=1}^r \,\,\sum_{ l_0+\ldots+l_d=2} :  \a^{(i)}_{l_0} \cdots \a^{(i)}_{l_d}:\,.
\end{equation}
Compare this with \eqref{Yboson}, it follows that
\begin{equation}
c(-2,d)=c(-1,1)\,c(-1,d) \times 2(d-1)=\sqrt{\beta}^{d+1}(d-1)/(d+1)\,.
\end{equation}
Similarly,
\ba
&&[D_{-1,0},D_{-1,d+1}]\sim
c(-1,0)\,c(-1,d+1) \times (d+2)\sum_{i=1}^r \,\,\sum_{ l_0+\ldots+l_d=2} :  \a^{(i)}_{l_0} \cdots \a^{(i)}_{l_d}:\nn\\
&&=\sqrt{\beta}^{d+1}
\sum_{i=1}^r \,\,\sum_{ l_0+\ldots+l_d=2} :  \a^{(i)}_{l_0} \cdots \a^{(i)}_{l_d}:\,.
\ea
Therefore, we have
\ba
&&D_{-2,d} = \left[D_{-1,0}\, , \,D_{-1,d+1} \right]-
\left[D_{-1,1} \, , \, D_{-1,d} \right]
\sim 2\sqrt{\beta}^{d+1}/(d+1)
\sum_{i=1}^r \,\,\sum_{ l_0+\ldots+l_d=2} :  \a^{(i)}_{l_0} \cdots \a^{(i)}_{l_d}:\nn\\
&&=\frac{2\sqrt{\beta}^{d+1}}{d+1} \big( :\! p_{d+1}(\underline{z})\!:\big)_{2}\,.
\ea
Some of the explicit expressions of  $W$-algebra in terms of SH$^c$ are given in the end of section \ref{sec:Whi-W}.
\section{Whittaker conditions in terms of SH$^c$}
\label{sec:Whi}
\subsection{$N_f$=0 case}
In order to prepare the generalization for $N_f\neq 0$, we present
the Whittaker condition for $N_f=0$
using our notation.
In the following, we demonstrate,
\ba
\label{eigen0}
D_{-1,d}|G\rangle=\k_d |G\rangle
\quad 0\leq d \leq N
\ea
with
\ba
\label{Kd0}
\k_d=\left\{
\begin{array}{ll}
0\quad& d<N-1\\
(-1)^{N-1}\frac{1}{\sqrt{\beta}}\L^{2} & d=N-1\\
(-1)^{N}\frac{1}{\sqrt{\beta}}\sum_{p}^N(a_p-\xi)\L^{2} & d=N \,.
\end{array} \right.
\ea

\paragraph{Proof:}
Set the coefficients in the Gaiotto state as,
\ba
u_{\vec W} := \Lambda^{2|\vec W|}(Z_\mathrm{vect}(\vec a, \vec W))\,.
\ea
Considering the action of SH operator given in \eqref{DW1} and \eqref{DW2}, one has
that for the Gaiotto state
\ba
D_{-1,d}|G\rangle&=&(-1)^{d} \sum_{\vec W}\sum_{q=1}^{N} \sum_{t=1}^{\tilde{f_q}}(a_q+B_t(W_q))^l  \Lambda^{(t,-)}_q(\vec W)
u_{\vec W}|\vec a,\vec W^{(t,-),q}\rangle.
\ea
If the Gaiotto state satisfies the Whittaker condition in \eqref{eigen0}, 
the following relation should hold:
\ba
\label{GWhi}
(-1)^{d} \sum_{\vec W (\supset \vec Y )}(a_q+B_t(W_q))^l  \Lambda^{(t,-)}_q(\vec W)
\frac{u_{\vec W}}{u_{\vec Y}}
=\k_d ,
\ea
where $\vec Y$ is obtained from $\vec W$ by removing one box:
$W_q^{(t,-)}=Y_q$, i.e., $W_q=Y_q^{(t,+)}$. We note that
$
\frac{\L^{2|\vec W|}}{\L^{2|\vec Y|}}
=\L^{2}\,.
$\\\\
For a Young diagram with one box removed or added (see Figure \ref{f:Young+-}), we find  $A_t(Y)$, $B_t(Y)$ (defined in \eqref{e:Ak} and \eqref{e:Bk}) in terms of their counterparts of the original Young diagram $W$:
\ba
\label{remove}
\begin{array}{ll}
A_t(W^{(k,-)})=\left\{
\begin{array}{ll}
A_t(W) & 1\leq t \leq k  \\
B_k(W) & t=k+1 \\
A_{t-1}(W) & k+2 \leq t \leq \tilde{f}+2 
\end{array} \right. \,,&
B_t(W^{(k,-)})=\left\{
\begin{array}{ll}
B_t(W) & 1\leq t \leq k-1  \\
B_k(W)-\beta & t=k \\
B_k(W)+1 & t=k+1 \\
B_{t-1}(W) & k+2 \leq t \leq \tilde{f}+1 
\end{array} \right.
\end{array} \,,
\ea

\ba
\label{add}
\begin{array}{ll}
A_s(W^{(k,+)})=\left\{
\begin{array}{ll}
A_s(W) & 1\leq s \leq k-1  \\
A_k(W)-1 & s=k \\
A_k(W)+\beta & s=k+1 \\
A_{s-1}(W) & k+2 \leq s \leq \tilde{f}+2 
\end{array} \right.\,,&
B_s(W^{(k,+)})=\left\{
\begin{array}{ll}
B_s(W) & 1\leq s \leq k-1  \\
A_k(W) & s=k \\
B_{s-1}(W) & k+1 \leq t \leq \tilde{f}+1 
\end{array} \right. 
\end{array} \,.
\ea
Using the above relations, after some lengthy computation referring to the appendix A.2 of  \cite{Kanno:2013aha}, 
we arrive at
\ba
&&\frac{u_{\vec W}}{u_{\vec Y}}
=\frac{u_{\vec Y^{(t,+),q}}}{u_{\vec Y}} \\
&&= \left(\frac{1}{\b}
\prod_{p=1}^N \left( \frac{{\prod}_{\ell=1}^{ f_p}(a_q-a_p+A_t(Y_q)-B_\ell(Y_p)+\xi)( a_q-a_p+A_t(Y_q)-B_\ell(Y_p))
}{{\prod}_{\ell=1}^{\prime f_p+1}( a_q-a_p+A_t(Y_q)-A_\ell(Y_p)-\xi) (a_q-a_p+A_t(Y_q)- A_\ell(Y_p))
}
\right)\right)^{1/2}\L^{2}\nn \,.
\ea
Therefore,
\ba
\k_d=(-1)^{d} \frac{1}{\sqrt{\beta}}\sum_{q=1}^{N} \sum_{t=1}^{\tilde{f_q}}(a_q+A_t(Y_q))^d 
\prod_{p=1}^N \left( \frac{{\prod}_{\ell=1}^{ f_p}(a_q-a_p+A_t(Y_q)-B_\ell(Y_p)+\xi)
}{{\prod}_{\ell=1}^{\prime f_p+1} (a_q-a_p+A_t(Y_q)- A_\ell(Y_p))
}
\right)\L^{2}\nn\,.
\ea
Setting
\ba
\begin{cases}
\label{xyset}
x_I=\{ a_p+A_k(Y_p)\} & 1\leq I\leq \sum_{p=1}^{N}( f_p+1)=\mathcal{N} \\
y_J=\{  a_p+B_\ell(Y_p)-\xi\} & 1\leq J\leq \sum_{p=1}^{N} f_p=\mathcal{M}
\end{cases}
\ea
where $\mathcal{N}-\mathcal{M}=N$,
we have $\k_d$ in a simplified form
\ba
\label{lambda}
\k_d=\L^{2}(-1)^{d} \frac{1}{\sqrt{\beta}}\sum_{I=1}^\mathcal{N} (x_I)^d \frac{\prod_{J=1}^\mathcal{M} (x_I-y_J)}{\prod_{J (\neq I)}^\mathcal{N} (x_I-x_J)}\,.
\ea
According to the formula used in \cite{Kanno:2013aha}:
\ba
\label{math}
\sum_{I=1}^{\mathcal{N}} (x_I)^m \frac{\prod_{J=1}^{\mathcal{M}} (x_I-y_J)}{\prod_{J (\neq I)}^{\mathcal{N}} (x_I-x_J)}
=\sum_{n=0}^{m+1+{\mathcal{M}}-{\mathcal{N}}} f_{m-n+1+{\mathcal{M}}-{\mathcal{N}}}(-y) b_{n}(x) ,
\ea
where
$
f_n(x)=\sum_{I_1<\cdots<I_n} x_{I_1}\cdots x_{I_n}
$ , 
and
$
b_n(x)=\sum_{I_1\leq\cdots \leq I_n }x_{I_1}\cdots x_{I_n}\,,$ 
we conclude that  $\k_d$ in \eqref{lambda} equals zero when  $d <N-1$, reduces to constant values in \eqref{Kd0} when $d=N-1, N$, but depends explicitly on $Y$ when $d >N$. 

\subsection{$N_f=k$ case}
In the following, we demonstrate that for $k<N$,
\ba
\uppercase\expandafter{\romannumeral1}: \qquad 
D_{-1,d}|G,m_1, \dots, m_k\rangle=\lambda_d |G,m_1, \dots, m_k\rangle
\quad 0\leq d \leq N-k
\ea
\ba
\uppercase\expandafter{\romannumeral2}: \qquad
D_{-2,d}|G,m_1, \dots, m_k\rangle=\lambda'_d |G,m_1, \dots, m_k\rangle
\quad 0\leq d \leq 2N-2k
\ea
with
\ba
\label{D1}
\lambda_d=\left\{
\begin{array}{ll}
0\quad&d<N-k-1\\
(-1)^{N-k-1}\frac{1}{\sqrt{\beta}}\L & d=N-k-1\\
(-1)^{N-k}\frac{1}{\sqrt{\beta}}\bigg(\sum_{p}^N(a_p-\xi) -\sum_{i=1}^k m_i \bigg)\L & d=N-k
\end{array}\right.
\ea
\ba
\label{D2}
\lambda'_d=\left\{
\begin{array}{ll}
0\quad& d<2N-2k-1\\
\L^{2} & d=2N-2k-1\\
-2\bigg(\sum_{p}^N(a_p-\xi) -\sum_{i=1}^k m_i \bigg)\L^{2} & d=2N-2k
\end{array}\right.
\ea
The above expressions still hold for $k=0$ case, but with the replacements $\L \to \L^2$.
Notice that $\lambda_{N-k+1}$ is not an eigenvalue but an operator which contains derivative of $\Lambda$:\\
 $\lambda_{N-k+1}=(-1)^{N-k+1}\frac{1}{\sqrt{\beta}}\bigg(
\b \L \frac{\partial}{\partial\L}+
\frac{1}{2}\sum_{p}^N(a_p-\xi)^2 +
\frac{1}{2}\big(\sum_{p}^N(a_p-\xi)\big )^2+
\sum_{i<j}^k m_im_j
-(\sum_{i=1}^k m_i) \sum_{p}^N(a_p-\xi) \bigg)\L$. \\We include this expression for later convenience.

\paragraph{Proof of \uppercase\expandafter{\romannumeral1}:}
Our proposal for the Gaiotto state takes the following form,
\ba
|G,m_1, \dots, m_k\rangle&=&\sum_{\vec Y}\L^{|\vec Y|}(Z_{\text {vect}})^{1/2}
\prod_{i=1}^{k} Z_{\text {fund}}(\vec a, \vec Y, m_i) |\vec a,\vec Y\rangle\,.
\ea
Since
\ba
\frac{Z_{\text {fund}}(\vec a, \vec Y^{(t,+),q}, m_1)}{Z_{\text {fund}}(\vec a, \vec Y, m_1)} 
&=&a_q+B_t(W_q)-m_1= a_q+A_t(Y_q)-m_1\nn ,
\ea
we find the action of $D_{-1,l}$ results to the similar form as
the one \eqref{GWhi} of  the $N_f=0$ case, and $\lambda_d$ is the generalized form of 
$\k_d$ in \eqref{lambda}:
\ba
\lambda_d=\L (-1)^{d} \frac{1}{\sqrt{\beta}}\sum_{I=1}^\mathcal{N} (x_I)^{d} \frac{\prod_{i=1}^{k} (x_I-m_i)\prod_{J=1}^\mathcal{M} (x_I-y_J)}{\prod_{J (\neq I)}^\mathcal{N} (x_I-x_J)}\,.
\ea
Again using \eqref{math}, we find that $\lambda_d$  reduces to \eqref{D1}.\\\\
\paragraph{Proof of \uppercase\expandafter{\romannumeral 2}:}
  To evaluate the action of $D_{-2,l}$ , we use the following commutation relations,
\ba
D_{-2,0} &=& \left[D_{-1,0}\, , \,D_{-1,1} \right]\\
D_{-2,1} &=& \left[D_{-1,0} \, , \, D_{-1,2} \right]\\
D_{-2,d} &=& \left[D_{-1,0}\, , \,D_{-1,d+1} \right]-
\left[D_{-1,1} \, , \, D_{-1,d} \right] \,.
\ea
Let us write the Gaiotto state as the following,
\ba
|G,m_1, \dots, m_k\rangle&=&\sum_{\vec W}c_{\vec W} |\vec a,\vec W\rangle,
\qquad
c_{\vec W} :=\L^{|\vec W|}(Z_{\text {vect}})^{1/2}
\prod_{i=1}^{k} Z_{\text {fund}}(\vec a, \vec Y, m_i) \,.
\ea
The action of $D_{-2,d}$ on the Gaiotto state is evaluated as
{\small
\ba
\label{D2d}
&&(-1)^{d+1} D_{-2,d }|G, m_1 \rangle \nonumber
\\  &&=
\sum_{q=1}^N\sum_{\ell=1}^{f_q} \beta \big((a_q+B_{\ell}(W_q))^d +(a_q+B_{\ell}(W_q) -\beta)^d \big) \Lambda_q^{(\ell,-2H)} (\vec W)c_{\vec W}|\vec a  ,\vec W^{(\ell,-2H),q}\rangle \nn
\\ &&\qquad \qquad \quad- \big((a_q+B_{\ell}(W_q))^d +(a_q+B_{\ell}(W_q) +1)^d \big)\Lambda_q^{(\ell,-2V)} (\vec W)c_{\vec W}|\vec a  ,\vec W^{(\ell,-2V),q}\rangle 
\\ &&-
\sum_{q=1}^N\sum_{u<\ell}^{f_q} 
\big((B_{u}(W_q)-B_{\ell}(W_q))\{(a_q+B_{u}(W_q))^d+(a_q+B_{\ell}(W_q))^d\}\big)
 \Lambda_q^{(\ell,-)} (\vec W) \Lambda_q^{(u,-)} (\vec W^{(\ell,-),q})c_{\vec W}|\vec a  ,\vec W^{(\ell,-;u,-),q}\rangle  \nonumber
\\ &&-
\sum_{q=1}^N\sum_{u<\ell}^{f_q} 
\big(B_{\ell}(W_q)-(B_{u}(W_q))\{(a_q+B_{u}(W_q))^d+(a_q+B_{\ell}(W_q))^d\}\big)
 \Lambda_q^{(u,-)} (\vec W) \Lambda_q^{(\ell+1,-)} (\vec W^{(u,-),q})
c_{\vec W}|\vec a  ,\vec W^{(\ell,-;u,-),q}\rangle  \nonumber
\\ &&=\lambda'_d\sum_{\vec Y}c_{\vec Y} |\vec a,\vec Y\rangle \nn
\ea}
where $Y^{(k,+2H)}$, $Y^{(k,+2V)}$ and $Y^{(k,+;u,+)}$ (resp. $Y^{(k,-2H)}$, $Y^{(k,-2V)}$
 and $Y^{(k,-;u,-)}$) stand for the Young diagrams obtained from adding (resp. deleting) two boxes horizontally, vertically and two different places, respectively.  \\$\Lambda_q^{(\ell,-2H)} $ etc. are defined
in \ref{WardL}. \\
The relations between $A_t(W)$, $B_t(W)$ and their counterparts of the original Young diagram $Y$ are
{\small
\ba
\begin{array}{ll}
A_k(W)=A_k(Y^{(l,+2H)})=\left\{
\begin{array}{ll}
A_k(Y) & 1\leq k \leq l-1  \\
A_l(Y)-1 & k=l \\
A_l(Y)+2\beta & k=l+1 \\
A_{k-1}(Y) & l+2 \leq k \leq \tilde{f}+2 
\end{array} \right.,&
B_k(Y)=B_k(Y^{(l,+2H)})=\left\{
\begin{array}{ll}
B_k(Y) & 1\leq k \leq l-1  \\
A_l(Y)+\beta & k=l \\
B_{k-1}(Y) & l+1 \leq k \leq \tilde{f}+1 
\end{array} \right. 
\end{array}.
\ea}
Again, after lengthy computations, we evaluate the four terms on the right hand side of \eqref{D2d} as below:
{\small
\ba
\Lambda_q^{(\ell,-2H)} (\vec W)\bigg(\frac{Z_{\text {vect}}(\vec W)}{Z_{\text {vect}}(\vec Y)}\bigg)^{1/2}
&=&
\frac{1}{\b(1+\b)}\sum_{I=1}^N  \frac{\prod_{I=1}^M (x_I-y_J)}{\prod_{J (\neq I)}^N (x_I-x_J)}
\frac{\prod_{I=1}^M (x_I-y_J+\b)}{\prod_{J (\neq I)}^N (x_I-x_J+\b)}
\, ,\\
\Lambda_q^{(\ell,-2V)} (\vec W)\bigg(\frac{Z_{\text {vect}}(\vec W)}{Z_{\text {vect}}(\vec Y)}\bigg)^{1/2}
&=&\frac{1}{1+\b}\sum_{I=1}^N  \frac{\prod_{I=1}^M (x_I-y_J)}{\prod_{J (\neq I)}^N (x_I-x_J)}
\frac{\prod_{I=1}^M (x_I-y_J-1)}{\prod_{J (\neq I)}^N (x_I-x_J-1)}
\, ,\nn\\
\Lambda_q^{(\ell,-)} (\vec W) \Lambda_q^{(u,-)} (\vec W^{(\ell,-),q})\bigg(\frac{Z_{\text {vect}}(\vec W)}{Z_{\text {vect}}(\vec Y)}\bigg)^{1/2}
&=&\frac{1}{2\b}\sum_{I=1}^{\mathcal{N}}
 \frac{\prod_{J=1}^{\mathcal{M}} (x_I-y_J)}
{\prod^{\mathcal{N}}_{J \neq I} (x_I-x_J)}
\sum_{K \neq I}^{\mathcal{N}}
 \frac{\prod_{J=1}^{\mathcal{M}} (x_K-y_J)}
{\prod^{\mathcal{N}}_{J \neq K} (x_K-x_J)}
\times \frac{(x_K-x_I)(x_K-x_I+1-\b)}{(x_K-x_I+1)(x_K-x_I-\b)}
\, ,\nn \\
 \Lambda_q^{(u,-)} (\vec W) \Lambda_q^{(\ell+1,-)} (\vec W^{(u,-),q})\bigg(\frac{Z_{\text {vect}}(\vec W)}{Z_{\text {vect}}(\vec Y)}\bigg)^{1/2}
&=&\frac{1}{2\b}\sum_{I=1}^{\mathcal{N}}
 \frac{\prod_{J=1}^{\mathcal{M}} (x_I-y_J)}
{\prod^{\mathcal{N}}_{J \neq I} (x_I-x_J)}
\sum_{K \neq I}^{\mathcal{N}}
 \frac{\prod_{J=1}^{\mathcal{M}} (x_K-y_J)}
{\prod^{\mathcal{N}}_{J \neq K} (x_K-x_J)}
\times \frac{(x_K-x_I)(x_K-x_I-1+\b)}{(x_K-x_I-1)(x_K-x_I+\b)}
\, ,\nn
\ea
}
where the redefinition of variables as in \eqref{xyset} are made.\\\\
As a result,  $\lambda'_d$ has the form,
{\small
\ba
&&(-1)^{d+1} \lambda'_d \\
&&=\frac{\L^{2}}{1+\b}\sum_{I=1}^\mathcal{N}  \frac{\prod_{I=1}^{\mathcal{M}}  (x_I-y_J)}{\prod_{J (\neq I)}^\mathcal{N} (x_I-x_J)}
\frac{\prod_{I=1}^{\mathcal{M}} (x_I-y_J+\b)}{\prod_{J (\neq I)}^\mathcal{N} (x_I-x_J+\b)} \times \big(x_I^d +(x_I+\b)^d\big) 
\times\prod_{i=1}^{k}\big( ( x_I-m_i)(x_I+\b-m_i)\big)   \nn\\
&&-\frac{\L^{2}}{1+\b}\sum_{I=1}^\mathcal{N}  \frac{\prod_{I=1}^{\mathcal{M}}  (x_I-y_J)}{\prod_{J (\neq I)}^\mathcal{N} (x_I-x_J)}
\frac{\prod_{I=1}^{\mathcal{M}}  (x_I-y_J-1)}{\prod_{J (\neq I)}^\mathcal{N} (x_I-x_J-1)}\times \big(x_I^d +(x_I-1)^d\big)
\times \prod_{i=1}^{k}\big( ( x_I-m_i)(x_I-1-m_i)\big)   \nn\\
&&+\frac{\L^{2}}{2\b}\sum_{I=1}^{\mathcal{N}}
 \frac{\prod_{J=1}^{\mathcal{M}} (x_I-y_J)}
{\prod^{\mathcal{N}}_{J \neq I} (x_I-x_J)}
\sum_{K \neq I}^{\mathcal{N}}
 \frac{\prod_{J=1}^{\mathcal{M}} (x_K-y_J)}
{\prod^{\mathcal{N}}_{J \neq K} (x_K-x_J)}
\times \frac{(x_K-x_I)^2(x_K-x_I+1-\b)}{(x_K-x_I+1)(x_K-x_I-\b)} \times \big(x_K^d + x_I^d\big)
\times\prod_{i=1}^{k}\big( ( x_K-m_i)( x_I-m_i)\big)\nn\\
&&-\frac{\L^{2}}{2\b}\sum_{I=1}^{\mathcal{N}}
 \frac{\prod_{J=1}^{\mathcal{M}} (x_I-y_J)}
{\prod^{\mathcal{N}}_{J \neq I} (x_I-x_J)}
\sum_{K \neq I}^{\mathcal{N}}
 \frac{\prod_{J=1}^{\mathcal{M}} (x_K-y_J)}
{\prod^{\mathcal{N}}_{J \neq K} (x_K-x_J)}
\times \frac{(x_K-x_I)^2(x_K-x_I-1+\b)}{(x_K-x_I-1)(x_K-x_I+\b)} \times \big(x_K^d + x_I^d\big)
\times \prod_{i=1}^{k}\big( ( x_K-m_i)( x_I-m_i)\big)
\, .\nn
\ea}
We note that a similar computation appears in the recursion formula with bifundamental multiplet  \eqref{l2sum}. After some algebra, it is simplified to
\ba
&&(-1)^{d+1} \lambda'_d  \\
&&=\frac{\L^{2}}{2(1+\b)}\sum_{I=1}^\mathcal{N}   \frac{\prod_{I=1}^{\mathcal{M}}  (x_I-y_J)}{\prod_{J (\neq I)}^\mathcal{N}  (x_I-x_J)}
\frac{\prod_{I=1}^{\mathcal{M}}  (x_I-y_J+\b)}{\prod_{J (\neq I)}^\mathcal{N}  (x_I-x_J+\b)} \times \big(x_I^d +(x_I+\b)^d\big) 
\times \prod_{i=1}^{k}\big( ( x_I-m_i)(x_I+\b-m_i)\big)   \nn\\
&&-\frac{\L^{2}}{2(1+\b)}\sum_{I=1}^\mathcal{N}   \frac{\prod_{I=1}^{\mathcal{M}}  (x_I-y_J)}{\prod_{J (\neq I)}^N (x_I-x_J)}
\frac{\prod_{I=1}^{\mathcal{M}}  (x_I-y_J-\b)}{\prod_{J (\neq I)}^\mathcal{N}  (x_I-x_J-\b)} \times \big(x_I^d +(x_I-\b)^d\big) 
\times\prod_{i=1}^{k}\big( ( x_I-m_i)(x_I-\b-m_i)\big)  \nn\\
&&-\frac{\L^{2}}{2(1+\b)}\sum_{I=1}^\mathcal{N}   \frac{\prod_{I=1}^{\mathcal{M}}  (x_I-y_J)}{\prod_{J (\neq I)}^N (x_I-x_J)}
\frac{\prod_{I=1}^{\mathcal{M}}  (x_I-y_J-1)}{\prod_{J (\neq I)}^\mathcal{N}  (x_I-x_J-1)}\times \big(x_I^d +(x_I-1)^d\big)
\times \prod_{i=1}^{k}\big( ( x_I-m_i)(x_I-1-m_i)\big)  \nn\\
&&+\frac{\L^{2}}{2(1+\b)}\sum_{I=1}^\mathcal{N}   \frac{\prod_{I=1}^{\mathcal{M}}  (x_I-y_J)}{\prod_{J (\neq I)}^\mathcal{N}  (x_I-x_J)}
\frac{\prod_{I=1}^{\mathcal{M}}  (x_I-y_J+1)}{\prod_{J (\neq I)}^\mathcal{N}  (x_I-x_J+1)}\times \big(x_I^d +(x_I+1)^d\big)
\times \prod_{i=1}^{k}\big( ( x_I-m_i)(x_I+1-m_i)\big) 
\, ,\nn
\ea
with $\mathcal{N}- \mathcal{M}=N$ .
In this form, one may use the trick (\ref{math}) to arrive at  \eqref{D2} .

\section{Whittaker conditions in terms of $W$-algebra}
\label{sec:Whi-W}
In this section, we rewrite the generalized Whittaker conditions obtained in the previous
section in terms of $W$-algebra
$W^{(d)}(z)=\sum_i W^{(d)}_{i} \, \, z^{-i-d}$.
Theorem 2 in the following is the main claim of the paper.
\begin{theorem}
{ For $N_f=0$ case \cite{r:SV},}
\ba
W^{(d)}_{1}|G\rangle=\lambda^{(d)}_{1} |G\rangle
\quad 0\leq d \leq N+1
\ea
with
\ba
\label{0lamd1}
\lambda^{(d)}_{1}=\left\{
\begin{array}{ll}
0\quad& d<N\\
(\sqrt{\beta})^{-N}\L^2 & d=N\\
(\sqrt{\beta})^{-N-1}\bigg(\frac{1}{N+1}\sum_{p}^N(a_p-\xi)
+ \frac{(N-1)N^2\xi}{2(N+1)} \bigg)\L^2
 & d=N+1
\end{array}\right. ,
\ea
and
\ba
W^{(d)}_{2}|G\rangle=0
\quad 0\leq d \leq 2N
\ea
Actually, for $SU(N)$ case we only have to consider up to $W^{(N)}$. From the commutation relations, it is obvious that
$W^{(d)}_{m}|G\rangle=0$ for $m\geq 2$ and $0\leq d \leq N$.
\end{theorem}
\begin{theorem}
{  For the Gaiotto state with $k$ fundamentals,}  one has
%
\ba
&& W^{(d)}_{1}|G,m_1, \dots, m_k\rangle=\lambda^{(d)}_{1}|G,m_1, \dots, m_k\rangle
\quad 0\leq d \leq N-k+1 ,\\
&& W^{(d)}_{2}|G,m_1, \dots, m_k\rangle=\lambda^{(d)}_{2} |G,m_1, \dots, m_k\rangle
\quad 0\leq d \leq 2N-2k+1 .
\ea
When $N-k>1$,
\ba
\label{lamd1}
\lambda^{(d)}_{1}=\left\{
\begin{array}{ll}
0\quad& d<N-k\\
(\sqrt{\beta})^{k-N}\L & d=N-k\\
(\sqrt{\beta})^{k-N-1}\bigg(\frac{1}{N-k+1}\sum_{p}^N(a_p-\xi) -\sum_{i=1}^k m_i
+ \frac{(N-k)(N-1)N\xi}{2(N-k+1)} \bigg)\L
 & d=N-k+1
\end{array}\right. ,
\ea
and
\ba
\label{lamd2}
\lambda^{(d)}_{2}=0
\quad& d<2N-2k+2 .
\ea
When $N-k=1$, 
\ba
&&  \lambda^{(1)}_{1}=-\frac{1}{\b}\L, \quad \lambda^{(2)}_{1}=\frac{1}{\b}\bigg(\sum_{p}^N(a_p-\xi) -\sum_{i=1}^k m_i \bigg)\L ,
\\
&& \lambda^{(2)}_{2}=\frac{1}{2\b}\L^{2}  ,\quad 
\lambda^{(3)}_{2}=\frac{1}{3\sqrt{\beta}\b}\bigg(\sum_{p}^N(a_p-\xi) -\sum_{i=1}^k m_i \bigg)\L^2 .
\ea
\end{theorem}
Before giving the proof of theorems, we give some comments.
\paragraph{Comments on the other generators:}
\begin{enumerate}
\item The action of $\lambda^{(N-k+2)}_{1}$ becomes an operator
involving the derivative of $\Lambda$ as we show later in \eqref{N-k+2}, and we see that 
 \ba
W^{(N-k+2)}_{1}|G, m_1,\cdots, m_k\rangle\sim (\frac{1}{\sqrt{\beta}} \L \frac{\partial}{\partial\L}+ const) 
\L |G, m_1,\cdots, m_k\rangle.
\ea
On the other hand, referring to \cite{Kanno:2013aha}  we have
{\small
\ba
&&J_0|G,m_1, \dots, m_k\rangle=\frac{1}{\b} \left(
-\sum_{p}^N (a_p-\xi) +\frac{\xi N(N-1)}{2} \right)|G,m_1, \dots, m_k\rangle,\label{eigenJ}\\
&&L_0|G,m_1, \dots, m_k\rangle\nonumber\\
&&~~~~~=\left( \L \frac{\partial}{\partial\L}+\frac{1}{2\beta} 
\left(
\sum_{p}^N(a_p-\xi)^2 +(1-N)\xi\sum_{p}^N(a_p-\xi) + \frac{\xi^2}{6} N(N-1)(N-2)
\right)\right)|G,m_1, \dots, m_k\rangle\,\label{eigenL}
\ea}
Compare to \eqref{eigenL}, we find in the action of $(W^{(N-k+2)}_{1}-\frac{1}{\sqrt{\beta}}\L L_0)$, the derivative of $\Lambda$ cancels. 
\item $W^{(d)}_{3}$ and higher can be generated by commutators of $W^{(r)}_{2}$, $W^{(r)}_{1}$ and $W^{(r)}_{0}$, with $r\leq d$, more precisely speaking, with the help of \eqref{yrd}. 
For example, when the action of both $L_{n-1}$ and $(W^{(N-k+2)}_{1}-\frac{1}{\sqrt{\beta}}\L L_0)$ 
on the Gaiotto state are constant, 
we have $W^{(3)}_{n}=\frac{1}{2n-3} [L_{n-1}, W^{(3)}_{1}]\sim\frac{1}{2n-3} [L_{n-1}, \frac{1}{\sqrt{\beta}}\L L_0]$, so
\ba
W^{(3)}_{n}|G,m_1, \dots, m_k\rangle=\frac{1}{2n-3} [L_{n-1}, \frac{1}{\sqrt{\beta}}\L L_0]|G,m_1, \dots, m_k\rangle=\frac{(n-1)\L}{(2n-3)\sqrt{\beta}}L_{n-1}|G,m_1, \dots, m_k\rangle
\ea
\end{enumerate}

\paragraph{Examples}
Here we give some simple cases 
of our theorem which match with the known results in the literature.
\begin{itemize}
\item{\bf $SU(2)$ case}
\ba
L_1|G\rangle&=&\frac{1}{\b}\L^2|G\rangle, \\
L_1|G,m\rangle&=&\frac{1}{\b}\bigg(\sum_{p}^2(a_p-\xi) -m\bigg)\L|G,m\rangle,\\ 
L_2|G,m\rangle&=&\frac{1}{2\b}\L^{2}|G,m\rangle. 
\ea
All higher $L_n$ have eigenvalue $0$. 
\item{\bf $SU(3)$ case}
\ba
L_1|G,m\rangle&=&\frac{1}{\b}\L|G,m\rangle\,, \\
W^{(3)}_1|G,m\rangle&=&\frac{1}{\sqrt{\beta}\b}\bigg(\frac{1}{3}\sum_{p}^3(a_p-\xi)
 -m+2\xi \bigg)\L^2|G,m\rangle\,, \\
L_1|G,m_1,m_2\rangle&=&\frac{1}{\b}\bigg(\sum_{p}^3(a_p-\xi) -(m_1+m_2) \bigg)\L|G,m_1,m_2\rangle\,, \\
L_2|G,m_1,m_2\rangle&=&\frac{1}{2\b}\L^{2}|G,m_1,m_2\rangle\,, 
\ea
\ba
W^{(3)}_{1}|G,m_1,  m_2\rangle
&=&\frac{1}{\sqrt{\beta}\beta}\bigg\{ \beta\L \frac{\partial}{\partial\L}+\frac{1}{2} 
\sum_{p}^3(a_p-\xi)^2 +\frac{1}{6} 
\big(\sum_{p}^3(a_p-\xi)\big)^2
+m_1m_2 \\
&+&2\xi(m_1+m_2) - \frac{1}{3}(m_1+m_2)\sum_{p}^3(a_p-\xi)
+ 3\xi^2
\bigg\}\L|G,m_1,  m_2\rangle\,, \nn \\
W^{(3)}_2|G,m_1,m_2\rangle&=&\frac{1}{3\sqrt{\beta}\b}\bigg(\sum_{p}^3(a_p-\xi)
 -(m_1+m_2) \bigg)\L^2|G,m_1,m_2\rangle\,, \\
W^{(3)}_3|G,m_1,m_2\rangle&=&\frac{1}{3\sqrt{\beta}\b}\L^3|G,m_1,m_2\rangle\,.
\ea
All higher $L_n$, $W_n$ have eigenvalue $0$. 
Since $\sum_{p}^N(a_p-\xi) $ can take arbitrary value, after set it to be zero we find the
above equations are in agreement with the known results\cite{r:Gaiotto, irregular, Kanno:2012},
up to overall constant coefficients. In order to compare with the result of \cite{Kanno:2012} , we have to remove the U(1) factor $\mathcal{J}(z)=\sum_{i=1}^n \partial\varphi_i(z)=:\! p_1(\underline{z})\!:$. Then we have $ L'_1=L_1-\frac{1}{N} \big( :\! p_{1}(\underline{z})\!:\big)_{0}
\big( :\! p_{1}(\underline{z})\!:\big)_{1}=L_1+\frac{1}{N}D_{-1,0}\sqrt{\beta}J_0$, and $ L'_2=L_2-\frac{1}{2N}
\big( :\! p_{1}(\underline{z})\!:\big)^2_{1}=L_2-\frac{1}{2N}(D_{-1,0})^2$ , thus
\ba
L'_1|G,m_1,m_2\rangle&=&\frac{1}{\b}\bigg(\frac{2}{3}\sum_{p}^3(a_p)-\xi -(m_1+m_2) \bigg)\L|G,m_1,m_2\rangle\,, \\
L'_2|G,m_1,m_2\rangle&=&\frac{1}{3\b}\L^{2}|G,m_1,m_2\rangle\,, 
\ea
which are consistent with those in \cite{Kanno:2012} by setting $\sum_{p}^N(a_p) =0 $ .
\end{itemize}
\paragraph{Proof of the theorems}

Up to terms of order $d-1$, the generators of W-algebra has the form
\begin{equation}
\label{Wrz}
W^{(d)}(z)\sim - \sum_{s=0}^d (-d)^{s-d} 
:\! p_1(\underline{z})^{d-s} e_s(\underline{z}) \!:
\end{equation}
where $e_l=\sum_{i_1<\cdots<i_l} z_{i_1}\cdots z_{i_l}$ is the 
elementary symmetric polynomial.
Then
using the expansion
\begin{equation}
\label{sym id}
e_n= -(-1)^{n}\frac{1}{n} p_n+\frac{1}{2}\sum_{r+s=n, r,s\geq 1}(-1)^{n}\frac{1}{rs}p_r p_s-\frac{1}{6}\sum_{r+s+t=n, r,s, t\geq 1}(-1)^{n}\frac{1}{rst}p_r p_s p_t
+\cdots,
\end{equation} 
it is deduced that, up to terms of order $d-1$,
\begin{equation}
W^{(d)}_{1}=(-1)^{d-1}(\sqrt{\beta})^{1-d} D_{-1,d-1} + u
\end{equation}
where $u$ is a linear combination of monomials 
$(D_{0,r_1} \cdots D_{0,r_s} D_{-1,r})$ with $r <d-1$, most of  which vanish when operate on the Gaiotto states. 
Take into consideration of \eqref{D1}, \eqref{D2}, we find explicit correspondence between
the generators.
In the following ``$\equiv$" means equivalent up to terms  which vanish when operate on the Gaiotto states).

Firstly for $W_1^{(d)}$ generators, 
\begin{itemize}
\item {For $N-k>1$},
\begin{equation}
\begin{split}
\label{N-k+2}
W^{(N-k+2)}_{1}\equiv&(-1)^{N-k+1}(\sqrt{\beta})^{k-N-1}D_{-1,N-k+1} 
-(-1)^{N-k+1}\frac{N-k+1}{N-k+2}(\sqrt{\beta})^{k-N+1}J_0 D_{-1,N-k}\\
&+(-1)^{N-k+1}\frac{(N-k+2)^2-2(N-k+2)-2}{2(N-k+2)^2}(\sqrt{\beta})^{k-N+3}J_0^2 D_{-1,(N-k-1)}\,,
\end{split}
\end{equation}
\begin{equation}
W^{(N-k+1)}_{1}\equiv(-1)^{N-k}(\sqrt{\beta})^{k-N}D_{-1,N-k} 
-(-1)^{N-k}\frac{N-k}{N-k+1}(\sqrt{\beta})^{2+k-N}J_0 D_{-1,N-k-1}\,,
\end{equation}
\begin{equation}
W^{(N-k)}_{1}\equiv(-1)^{N-k-1}(\sqrt{\beta})^{1+k-N} D_{-1,N-k-1}\,.
\end{equation}
\item  For $N-k=1$, 
\be
W^{(3)}_{1}\equiv\frac{1}{\b}D_{-1,2} -\frac{2}{3}J_0 D_{-1,1}+
\frac{1}{3}\b J_0^2 D_{-1,0}\,,
\ee
\be
W^{(2)}_{1}=L_1\equiv(-\sqrt{\beta})^{-1} D_{-1,1}\,,
\ee
\be
W^{(1)}_{1}=J_1\equiv(-\sqrt{\beta})^{-1} D_{-1,0}\,.
\ee
\end{itemize}
%
Secondly for $W^{(d)}_2$ generators are related to SH as,
\begin{equation}
\label{Wr2}
W^{(d)}_{2}=\frac{1}{2\sqrt{\beta}^{d}}(-1)^{d} D_{-2,d-1} + u'
\end{equation}
This time $u'$ is a linear combination of monomials 
$(D_{0,r_1} \cdots D_{0,r_s} D_{-1,r}D_{-2,r})$ with $r <d-1$, again most of which vanish when operate on the Gaiotto states.  Explicitly,
\begin{itemize}
\item For $N-k>1$, 
\begin{equation}
\begin{split}
W^{(2N-2k+1)}_{2}&\equiv-\frac{1}{2\sqrt{\beta}\b^{N-k}} D_{-2, 2N-2k} +\frac{N-k}{2N-2k+1}\frac{\sqrt{\beta}}{\b^{N-k}} J_0 D_{-2,2N-2k-1}\\
&+\frac{1}{\sqrt{\beta}\b^{N-k-1}} D_{-1,N-k-1}D_{-1,N-k}
-\frac{N-k}{2N-2k+1}\frac{\sqrt{\beta}}{\b^{N-k-1}} J_0 (D_{-1,N-k-1})^2,
\end{split}
\end{equation}
\begin{equation}
W^{(2N-2k)}_{2}\equiv\frac{1}{2\b^{N-k}} D_{-2,2N-2k-1} 
-\frac{1}{2\b^{N-k-1}}(D_{-1,N-k-1})^2 .
\end{equation}
\item For $N-k=1$, 
\begin{equation}
W^{(3)}_{2}\equiv-\frac{1}{2\sqrt{\beta}\b} D_{-2, 2}+\frac{2}{3\sqrt{\beta}} D_{-1,0}D_{-1,1} +\frac{1}{3\sqrt{\beta}} J_0 D_{-2,1}
-\frac{\sqrt{\beta}}{3} J_0 (D_{-1,0})^2,
\end{equation}
\be
W^{(2)}_{2}=L_2\equiv\frac{1}{2\b} D_{-2,1} \,.
\ee
\end{itemize}
Combining with \eqref{Kd0}, \eqref{D1}and \eqref{D2}, the above equations lead straightforwardly to \eqref{0lamd1}, \eqref{lamd1} and \eqref{lamd2}  in the beginning of this section. 
\section{Conclusion}
Inspired by AGT conjecture, we construct Gaiotto states with fundamental
multiplets in $SU(N)$ gauge theories by splitting the corresponding Nekrasov partition function in a proper 
way, and prove that they satisfy the requirements of Whittaker vectors. We make use of a useful algebra SH. Though SH is complicated in form, it has nice properties when acts on the Hilbert space. Also by clarifying its relation with $W_n$ algebra, we are able to obtain the eigenvalues of higher spin $W_n$ generators for general SU$(N)$ case, extending the current methods limited to SU$(3)$.
For the future work we will construct Gaiotto states for linear quiver theory, and compare with another type of Gaiotto state arising from the colliding limit \cite{GT2012, KMST2013}. 
In this way, it would be interesting to find the explicit connection between this result and the 
coherent state approach found in \cite{RNC}.
\\\\
As another application of SH we complete the discussion of Virasoro constraint for Nekrasov partition function's recursion relation, by calculating the $L_{\pm 2}$ constraints directly. Combined with the  $J_{\pm 1}$and $L_{\pm 1}$ constraints showed in  \cite{Kanno:2013aha}, this non-trivial relation gives a strong support for SU$(N)$ AGT conjecture of  linear quiver type. 
Especially for SU$(2)$ case, Virasoro constraint is enough to serve as a proof of AGT conjecture.  An interesting extension to W algebra constraint is now made more accessible since we can easily write down the explicit relation between SH and $W_n$ algebra.

\subsection*{Acknowledgments}
YM thanks Hiroshi Itoyama, Hiroaki Kanno and Yasuhiko Yamada for
the discussion on DAHA and Gaiotto states. YM is supported in part by KAKENHI (\#25400246). HZ thanks the former members of the particle physics group in Chuo University 
for helpful discussions, and owes special thanks to Takeo Inami for his instructions and kind support. This work is partially supported by the National Research Foundation of Korea (NRF)  (NRF-2013K1A3A1A39073412) (CR), and (NRF-2014R1A2A2A01004951) (CR and HZ). 
\appendix
\section{Derivation of $L_{\pm 2}$ constraints on the bifundamental multiplets}
In this appendix, we derive a proof of Ward identities for $L_{\pm 2}$ which was not given in \cite{Kanno:2013aha}.  While this is extremely technical, it is important to show the Nekrasov partition function for the bifundamental matter has the invariance with respect to Virasoro generators $L_n$. 
This section in general follows the same construction as \cite{Kanno:2013aha}. \\\\
The instanton partition function for linear quiver gauge theories
is decomposed into matrix like product with a factor  $Z_{\vec Y, \vec W}$
which depends on two sets of Young diagrams.
Here the Young diagrams $\vec Y=(Y_1,\cdots, Y_N)$ 
represent the fixed points of
$U(N)$ instanton moduli space under localization.
$Z_{\vec Y,\vec W}$ consists of contributions from one bifundamental 
hypermultiplet and vectormultiplets.
We find that the building block $Z_{\vec Y, \vec W}$ 
satisfies an infinite series of recursion relations,
\ba \label{e:sketch}
\delta_{\pm m,n} Z_{\vec Y,\vec W} -U_{\pm m,n} Z_{\vec Y,\vec W}=0\,,
\ea
where $\delta_{\pm m,n}Z_{\vec Y,\vec W}$  
represents a sum of the Nekrasov partition function with 
instanton number larger  or less  than $Z_{\vec Y,\vec W}$ by $m$
with appropriate coefficients, and $U_{\pm m,n}$ are polynomials
of parameters such as the mass of bifundamental matter or the VEV of gauge multilets.
The subscript $m$ takes arbitrary integer values and $n$ takes any non-negative integer values.
We observe that AGT conjecture can be proved once we prove the relation
\ba\label{conj}
Z(\vec a, \vec Y; \vec b, \vec W;\mu)=\langle \vec a + \nu \vec e, \vec Y|V(1) |\vec b + (\xi + \nu+\mu) \vec e, \vec W\rangle,
\ea
\subsection{Modified vertex operator for $U(1)$ factor}
The free boson field which describes the $U(1)$ part is given 
by the operators $J_n$ defined in the previous section.
We modify the vertex operator $\tilde V^H$ for the $U(1)$ factor as,
\ba
&& V^H_\kappa(z) =e^{\frac{1}{\sqrt N}(NQ-\kappa) \phi_-}
 e^{\frac{-1}{\sqrt N}\kappa \phi_+}\,,\\
&& \phi_+=\alpha_0 \log z -\sum_{n=1}^\infty \frac{\alpha_n}{n} z^{-n}\,,\quad
\phi_-=q +\sum_{n=1}^\infty \frac{\alpha_{-n}}{n} z^{n}\,.
\ea

The general commutator $\left[L_n, V_\kappa(z)\right]$ is given in \cite{Kanno:2013aha} , here 
we write the special cases $n=\pm2$ for the convenience of later calculation.
\ba 
&&\left[L_2, V_\kappa(z)\right] \\ \nn
&&=z^{3}\partial_z V_\kappa(z)
+\frac{3(NQ-\kappa)^2}{2N}z^{2} V_\kappa(z)
+\sqrt{N} Q z^{2} V_\kappa(z) \alpha_0
+\sqrt{N} Q z V_\kappa(z) \alpha_1
+\sqrt{N} Q  V_\kappa(z) \alpha_2
+3z^{2}\Delta_W V_\kappa(z) \,,
\ea
\ba 
\left[L_{-2}, V_\kappa(z)\right]&=&z^{-1}\partial_z V_\kappa(z)
-\frac{\kappa^2}{2N}z^{-2}  V_\kappa(z)
-\sqrt{N} Q z^{-1}\alpha_{-1} V_\kappa(z)
-z^{-2}\Delta_W V_\kappa(z)\,.
\ea
where 
$
\Delta_W=\frac{\kappa(\kappa-Q(N-1))}{2}-\frac{\kappa^2}{2N}
$ is the conformal dimension of $W_N$ vertex operator $V^W_\kappa$ with Toda momenta 
$\vec p=-\kappa(\vec e_N-\frac{\vec e}{N})$.

\subsection{Ward identities for $J_{\pm 1}$ and $L_{\pm 1}$}
These analysis have already been performed in \cite{Kanno:2013aha} , and we obtained the
following:\\
The Ward identity for $J_1$
is proved since it is identified with the recursion formula 
$\delta_{-1,0} Z_{\vec Y,\vec W} -U_{-1,0} Z_{\vec Y,\vec W}=0$.\\
It shows the equivalence between 
the recursion formula $\delta_{1,0} Z_{\vec Y,\vec W} -U_{1,0} Z_{\vec Y,\vec W}=0$
and the Ward identity for $J_{-1}$.\\
The Ward identity for $L_1$ is reduced to
the recursion relation $\delta_{-1,1} Z_{\vec Y,\vec W} -U_{-1,1} Z_{\vec Y,\vec W}=0$.
In the same way, for $L_{-1}$,  the recursion formula  $\delta_{1,1} Z_{\vec Y,\vec W} -U_{1,1} Z_{\vec Y,\vec W}=0$ can be identified with the Ward identity.
These consistency conditions are highly nontrivial
and strongly suggest that
the identify (\ref{e:sketch}) are a part of
the Ward identities for the extended conformal symmetry.


\subsection{Ward identities for $L_{\pm 2}$}
\label{WardL}
Our goal is to show the recursion formula 
$\delta_{\pm2,1} Z_{\vec Y,\vec W} -U_{\pm2,1} Z_{\vec Y,\vec W}=0$.
From the definition of $L_n$\eqref{defV},
\begin{equation}
\begin{split}
 L_2=\frac{(-\sqrt{\beta})^{-2}}{2} D_{-2,1} - \frac{N \xi}{2} J_2 =\frac{1}{2\beta} [D_{-1,0},D_{-1,2}] -\frac{1}{2\beta}N\xi [D_{-1,0},D_{-1,1}]
\end{split}
\end{equation}
The action of the commutator on the basis reads,
\ba
&&\langle \vec a + \nu \vec e, \vec Y| \frac{1}{\beta} [D_{-1,0},D_{-1,2}]\nonumber
\\ &&=\frac{1}{\beta}\sum_{p=1}^N\sum_{k=1}^{f_p}
\langle \vec a + \nu \vec e,\vec  Y^{(k,+2H),p}|
 \beta( 2a_p+ 2\nu+2A_{k}(Y_p)+\beta) \Lambda_p^{(k,+2H)} (\vec Y) \nn
\\ &&\qquad \qquad \quad -\langle \vec a + \nu \vec e,\vec Y^{(k,+2V),p}| ( 2a_p+ 2\nu+2A_{k}(Y_p)-1)\Lambda_p^{(k,+2V)} (\vec Y)
\\ &&+
 \frac{-1}{\beta}\sum_{p=1}^N\sum_{u<k}^{f_p+1}
\langle \vec a + \nu \vec e,\vec  Y^{(k,+;u,+),p}|
 \Lambda_p^{(k,+)} (\vec Y)
  \Lambda_p^{(u,+)} (\vec Y^{(k,+),p})
  \bigg((A_{u}(Y_p)-A_{k}(Y_p))( 2a_p+ 2\nu+A_{k}(Y_p)+A_{u}(Y_p))
\bigg)\nonumber 
\\ &&+ \frac{-1}{\beta}\sum_{p=1}^N\sum_{u<k}^{f_p+1}
\langle \vec a + \nu \vec e,\vec  Y^{(k,+;u,+),p}|
 \Lambda_p^{(u,+)} (\vec Y)
  \Lambda_p^{(k+1,+)} (\vec Y^{(u,+),p})
  \bigg(A_{k}((Y_p)-A_{u}(Y_p)( 2a_p+ 2\nu+A_{k}(Y_p)+A_{u}(Y_p))
\bigg)\nonumber  
\ea
{\small
\ba
&&\frac{1}{\beta} [D_{-1,0},D_{-1,2}] |\vec b + (\xi + \nu+\mu) \vec e ,\vec W \rangle \nonumber
\\ &&=
\frac{1}{\beta}\sum_{q=1}^N\sum_{\ell=1}^{f_q} \beta(2b_q+ 2\nu+2\mu+2B_{\ell}(W_q)+2\xi -\beta) \Lambda_q^{(\ell,-2H)} (\vec W)|\vec b + (\xi + \nu+\mu) \vec e ,\vec W^{(\ell,-2H),q}\rangle \nn
\\ &&\qquad \qquad \quad-(2b_q+ 2\nu+2\mu+2B_{\ell}(W_q)+2\xi +1)\Lambda_q^{(\ell,-2V)} (\vec W)|\vec b + (\xi + \nu+\mu) \vec e ,\vec W^{(\ell,-2V),q}\rangle 
\\ &&-
\frac{1}{\beta}\sum_{q=1}^N\sum_{u<\ell}^{f_q} 
\big((B_{u}(W_q)-B_{\ell}(W_q))(2b_q+ 2\nu+2\mu+B_{u}(W_q)+B_{\ell}(W_q))\big)
 \Lambda_q^{(\ell,-)} (\vec W) \Lambda_q^{(u,-)} (\vec W^{(\ell,-),q})|\vec b + (\xi + \nu+\mu) \vec e ,\vec W^{(\ell,-;u,-),q}\rangle  \nonumber
\\ &&-
\frac{1}{\beta}\sum_{q=1}^N\sum_{u<\ell}^{f_q} 
\big(B_{\ell}(W_q)-(B_{u}(W_q))(2b_q+ 2\nu+2\mu+B_{u}(W_q)+B_{\ell}(W_q))\big)
 \Lambda_q^{(u,-)} (\vec W) \Lambda_q^{(\ell+1,-)} (\vec W^{(u,-),q})|\vec b + (\xi + \nu+\mu) \vec e ,\vec W^{(\ell,-;u,-),q}\rangle  \nonumber
\ea
}
In the two above equations, we have used the relation \eqref{remove} and \eqref{add} , and
\ba
\Lambda_q^{(\ell,-2H)} (\vec W) &=&\Biggl\{
\frac{2}{\beta+1}\prod_{p=1}^N \Biggl(\prod_{k=1}^{\tilde f_p+1} \frac{
 (b_q-b_p+B_l(W_q)-A_k(W_p)-\xi)(b_q-b_p+B_l(W_q)-A_k(W_p)-\xi-\beta)
}{
 (b_q-b_p+B_l(W_q)-A_k(W_p))(b_q-b_p+B_l(W_q)-A_k(W_p)-\beta)
} \nn \\
&&{\prod}_{k=1}^{\prime \tilde f_p}\frac{ (b_q-b_p+B_l(W_q)-B_k(W_p)+\xi)(b_q-b_p+B_l(W_q)-B_k(W_p)+\xi-\beta)}{
 ( b_q-b_p+B_l(W_q)-B_k(W_p))( b_q-b_p+B_l(W_q)-B_k(W_p)-\beta)
}
\Biggl)\Biggl\}^{1/2} \\
\Lambda^{(\ell,-2V),q} (\vec W) &=&\Biggl\{
\frac{2\beta}{\beta+1}\prod_{p=1}^N \Biggl(\prod_{k=1}^{\tilde f_p+1} \frac{
 (b_q-b_p+B_l(W_q)-A_k(W_p)-\xi)(b_q-b_p+B_l(W_q)-A_k(W_p)-\xi+1)
}{
 (b_q-b_p+B_l(W_q)-A_k(W_p))(b_q-b_p+B_l(W_q)-A_k(W_p)+1)
} \nn \\
&&{\prod}_{k=1}^{\prime \tilde f_p}\frac{ (b_q-b_p+B_l(W_q)-B_k(W_p)+\xi)(b_q-b_p+B_l(W_q)-B_k(W_p)+\xi+1)}{
 ( b_q-b_p+B_l(W_q)-B_k(W_p))( b_q-b_p+B_l(W_q)-B_k(W_p)+1)
}
\Biggl)\Biggl\}^{1/2}.
\ea
\ba
\Lambda^{(k,+2H),p} (\vec Y) &=&\Biggl\{
\frac{2}{\beta+1}\prod_{q=1}^N \Biggl(\prod_{\ell=1}^{f_q}  \frac{
 (a_p-a_q+A_k(Y_p)-B_\ell(Y_q)+\xi)(a_p-a_q+A_k(Y_p)-B_\ell(Y_q)+\xi +\beta)
}{
 (a_p-a_q+A_k(Y_p)- B_\ell(Y_q))(a_p-a_q+A_k(Y_p)- B_\ell(Y_q)+\beta)
} \nn \\
&&{\prod}_{\ell=1}^{\prime \tilde f_q}\frac{(a_p-a_q+A_k(Y_p)- A_\ell(Y_q) -\xi)(a_p-a_q+A_k(Y_p)- A_\ell(Y_q) -\xi+\beta)}{(a_p-a_q+A_k(Y_p)-A_\ell(Y_q))(a_p-a_q+A_k(Y_p)-A_\ell(Y_q)+\beta)}
\Biggl)\Biggl\}^{1/2} \\
\Lambda^{(k,+2V),p} (\vec Y) &=&\Biggl\{
\frac{2\beta}{\beta+1}\prod_{p=1}^N \Biggl(\prod_{k=1}^{\tilde f_p+1} \frac{
 (a_p-a_q+A_k(Y_p)-B_\ell(Y_q)+\xi)(a_p-a_q+A_k(Y_p)-B_\ell(Y_q)+\xi -1)
}{
 (a_p-a_q+A_k(Y_p)- B_\ell(Y_q))(a_p-a_q+A_k(Y_p)- B_\ell(Y_q)-1)
} \nn \\
&&{\prod}_{k=1}^{\prime \tilde f_p}\frac{(a_p-a_q+A_k(Y_p)- A_\ell(Y_q) -\xi)(a_p-a_q+A_k(Y_p)- A_\ell(Y_q) -\xi-1)}{(a_p-a_q+A_k(Y_p)-A_\ell(Y_q))(a_p-a_q+A_k(Y_p)-A_\ell(Y_q)-1)}
\Biggl)\Biggl\}^{1/2}  .
\ea
For $u<k$,
{\small
\ba
&&  \Lambda_p^{(u,+)} (\vec Y^{(k,+),p}) \nn \\
&&=\Lambda_p^{(u,+)} (\vec Y)
\times \frac{A_u(Y_p)-A_k(Y_p)+\xi}   {A_u(Y_p)-A_k(Y_p)}
\times \frac{A_u(Y_p)-A_k(Y_p)+\b}   {A_u(Y_p)-A_k(Y_p)+1}
\times \frac{A_u(Y_p)-A_k(Y_p)-1}   {A_u(Y_p)-A_k(Y_p)-\b}
\times \frac{A_u(Y_p)-A_k(Y_p)}   {A_u(Y_p)-A_k(Y_p)-\xi}
 \ea
 \ba
&&  \Lambda_p^{(k+1,+)} (\vec Y^{(u,+),p}) \nn \\
&&=\Lambda_p^{(k,+)} (\vec Y)
\times \frac{A_k(Y_p)-A_u(Y_p)+\xi}   {A_k(Y_p)-A_u(Y_p)}
\times \frac{A_k(Y_p)-A_u(Y_p)+\b}   {A_k(Y_p)-A_u(Y_p)+1}
\times \frac{A_k(Y_p)-A_u(Y_p)-1}   {A_k(Y_p)-A_u(Y_p)-\b}
\times \frac{A_k(Y_p)-A_u(Y_p)}   {A_k(Y_p)-A_u(Y_p)-\xi}
 \ea}
For convenience, we set (this convention is only used in this appendix, different from \eqref{xyset})\\
$x_I=\begin{cases}
\{ a_p+\nu+A_k(Y_p)\} & 1\leq I\leq \mathcal{N} \\
\{  b_p+\nu+\mu+B_k(W_p)\} & \mathcal{N}+1\leq I\leq \mathcal{N}+\mathcal{M}
\end{cases}$

$y_I=\begin{cases}
\{ a_p+\nu+B_k(Y_p)-\xi\} & 1\leq I\leq \mathcal{N}-N \\
\{b_p+\nu+\mu+A_k(W_p)+\xi \} & \mathcal{N}-N+1\leq I\leq \mathcal{N}+\mathcal{M} .
\end{cases}$\\\\
Like the $L_1$ case performed in \cite{Kanno:2013aha} , anomalous terms arise both from the 
the action on the ket basis and the modified vertex operator, and again cancels with each other:
 $2\xi$ terms exactly cancel with the contribution of $\sqrt{N} Q  V_\kappa(z) \alpha_2$, and the rest has the following form
\begin{equation}
\begin{split}
\label{l2sum}
&\frac{\langle \vec a + \nu \vec e
, \vec Y| L_2  V_\kappa(1) |\vec b + (\xi + \nu+\mu) \vec e
, \vec W\rangle}{\langle\vec a + \nu \vec e
, \vec Y| V_\kappa(1)|\vec b + (\xi + \nu+\mu) \vec e
, \vec W\rangle} 
-
\frac{\langle \vec a + \nu \vec e
, \vec Y|  V_\kappa(1) L_2|\vec b + (\xi + \nu+\mu) \vec e
, \vec W\rangle}{\langle\vec a + \nu \vec e
, \vec Y| V_\kappa(1)|\vec b + (\xi + \nu+\mu) \vec e
, \vec W\rangle} 
\\
&+\frac{N \xi}{2}
\frac{\langle \vec a + \nu \vec e
, \vec Y| [J_2 , V_\kappa(1)] |\vec b + (\xi + \nu+\mu) \vec e
, \vec W\rangle}{\langle\vec a + \nu \vec e
, \vec Y| V_\kappa(1)|\vec b + (\xi + \nu+\mu) \vec e
, \vec W\rangle} 
\\
&-{\sqrt \b}Q\frac{\langle \vec a + \nu \vec e
, \vec Y|   V_\kappa(1)J_1 |\vec b + (\xi + \nu+\mu) \vec e
, \vec W\rangle}{\langle\vec a + \nu \vec e
, \vec Y| V_\kappa(1)|\vec b + (\xi + \nu+\mu) \vec e
, \vec W\rangle} 
-{\sqrt \b}Q\frac{\langle \vec a + \nu \vec e
, \vec Y|   V_\kappa(1)J_2 |\vec b + (\xi + \nu+\mu) \vec e
, \vec W\rangle}{\langle\vec a + \nu \vec e
, \vec Y| V_\kappa(1)|\vec b + (\xi + \nu+\mu) \vec e
, \vec W\rangle} 
\\
&=-\frac{1}{2\b(\b+1)}\sum_{I=1}^{\mathcal{N}}
 \frac{\prod_{J=1}^{\mathcal{N}+\mathcal{M}} (x_I-y_J)}
{\prod^{\mathcal{N}+\mathcal{M}}_{J \neq I} (x_I-x_J)}
\frac{\prod_{J=1}^{\mathcal{N}+\mathcal{M}} (x_I-y_J-1)}
{\prod^{\mathcal{N}+\mathcal{M}}_{J \neq I} (x_I-x_J-1)}\times (2x_I-1)\\
&-\frac{1}{2\b(\b+1)}\sum_{I=\mathcal{N}+1}^{\mathcal{N}+\mathcal{M}}
 \frac{\prod_{J=1}^{\mathcal{N}+\mathcal{M}} (x_I-y_J)}
{\prod^{\mathcal{N}+\mathcal{M}}_{J \neq I} (x_I-x_J)}
\frac{\prod_{J=1}^{\mathcal{N}+\mathcal{M}} (x_I-y_J-\b)}
{\prod^{\mathcal{N}+\mathcal{M}}_{J \neq I} (x_I-x_J-\b)}\times (2x_I-\b)\\
&+\frac{1}{2\b(\b+1)}\sum_{I=1}^{\mathcal{N}}
 \frac{\prod_{J=1}^{\mathcal{N}+\mathcal{M}} (x_I-y_J)}
{\prod^{\mathcal{N}+\mathcal{M}}_{J \neq I} (x_I-x_J)}
\frac{\prod_{J=1}^{\mathcal{N}+\mathcal{M}} (x_I-y_J+\b)}
{\prod^{\mathcal{N}+\mathcal{M}}_{J \neq I} (x_I-x_J+\b)}\times (2x_I+\b)\\
&+\frac{1}{2\b(\b+1)}\sum_{I=\mathcal{N}+1}^{\mathcal{N}+\mathcal{M}}
 \frac{\prod_{J=1}^{\mathcal{N}+\mathcal{M}} (x_I-y_J)}
{\prod^{\mathcal{N}+\mathcal{M}}_{J \neq I} (x_I-x_J)}
\frac{\prod_{J=1}^{\mathcal{N}+\mathcal{M}} (x_I-y_J+1)}
{\prod^{\mathcal{N}+\mathcal{M}}_{J \neq I} (x_I-x_J+1)}\times (2x_I+1)\\
&+\frac{1}{4\b^2}\sum_{I=1}^{\mathcal{N}}
 \frac{\prod_{J=1}^{\mathcal{N}+\mathcal{M}} (x_I-y_J)}
{\prod^{\mathcal{N}+\mathcal{M}}_{J \neq I} (x_I-x_J)}
\sum_{K \neq I}^{\mathcal{N}}
 \frac{\prod_{J=1}^{\mathcal{N}+\mathcal{M}} (x_K-y_J)}
{\prod^{\mathcal{N}+\mathcal{M}}_{J \neq K} (x_K-x_J)}
\times \frac{(x_K-x_I)^2(x_K-x_I+1-\b)}{(x_K-x_I+1)(x_K-x_I-\b)}\times(x_K+x_I)\\
&-\frac{1}{4\b^2}\sum_{I=1}^{\mathcal{N}}
 \frac{\prod_{J=1}^{\mathcal{N}+\mathcal{M}} (x_I-y_J)}
{\prod^{\mathcal{N}+\mathcal{M}}_{J \neq I} (x_I-x_J)}
\sum_{K \neq I}^{\mathcal{N}}
 \frac{\prod_{J=1}^{\mathcal{N}+\mathcal{M}} (x_K-y_J)}
{\prod^{\mathcal{N}+\mathcal{M}}_{J \neq K} (x_K-x_J)}
\times \frac{(x_K-x_I)^2(x_K-x_I-1+\b)}{(x_K-x_I-1)(x_K-x_I+\b)}\times(x_K+x_I)\\
&-\frac{1}{4\b^2} \sum_{I=\mathcal{N}+1}^{\mathcal{N}+\mathcal{M}}
 \frac{\prod_{J=1}^{\mathcal{N}+\mathcal{M}} (x_I-y_J)}
{\prod^{\mathcal{N}+\mathcal{M}}_{J \neq I} (x_I-x_J)}
\sum_{K=\mathcal{N}+1,K \neq I}^{\mathcal{N}+\mathcal{M}}
 \frac{\prod_{J=1}^{\mathcal{N}+\mathcal{M}} (x_K-y_J)}
{\prod^{\mathcal{N}+\mathcal{M}}_{J \neq K} (x_K-x_J)}
\times \frac{(x_K-x_I)^2(x_K-x_I+1-\b)}{(x_K-x_I+1)(x_K-x_I-\b)}\times(x_K+x_I)\\
&+\frac{1}{4\b^2} \sum_{I=\mathcal{N}+1}^{\mathcal{N}+\mathcal{M}}
 \frac{\prod_{J=1}^{\mathcal{N}+\mathcal{M}} (x_I-y_J)}
{\prod^{\mathcal{N}+\mathcal{M}}_{J \neq I} (x_I-x_J)}
\sum_{K=\mathcal{N}+1,K \neq I}^{\mathcal{N}+\mathcal{M}}
 \frac{\prod_{J=1}^{\mathcal{N}+\mathcal{M}} (x_K-y_J)}
{\prod^{\mathcal{N}+\mathcal{M}}_{J \neq K} (x_K-x_J)}
\times \frac{(x_K-x_I)^2(x_K-x_I-1+\b)}{(x_K-x_I-1)(x_K-x_I+\b)}\times(x_K+x_I)\\
&-\frac{1-\b}{\b}\sum_{I=\mathcal{N}+1}^{\mathcal{N}+\mathcal{M}}
 \frac{\prod_{J=1}^{\mathcal{N}+\mathcal{M}} (x_I-y_J)}
{\prod^{\mathcal{N}+\mathcal{M}}_{J \neq I} (x_I-x_J)}.
\end{split}
\end{equation}
Using some tricks like redefining  $x'_I=x_1,x_2, \dots, x_{I-1} , x_{I+1},\dots, x_{\mathcal{N}+\mathcal{M}} $ plus $x_I-1, x_I+\b$,  
and $y'_I=y_1,y_2, \dots, y_{\mathcal{N}+\mathcal{M}} $ plus $x_I-1+\b$, the above
can be evaluated by \eqref{math}, and finally reduces to
\ba
\sqrt{\beta}^{-1} \frac{\delta_{-2,1} Z(\vec a, \vec Y; \vec b, \vec W;\mu)}
{ Z(\vec a, \vec Y; \vec b, \vec W;\mu)}
+\sqrt {\beta}^{-1}\frac{N \xi}{2} \frac{\delta_{-2,0} Z(\vec a, \vec Y; \vec b, \vec W;\mu)}
{ Z(\vec a, \vec Y; \vec b, \vec W;\mu)}\, .\nn
\ea On the other hand, the commutator part becomes
\begin{equation}
\begin{split}
 \label{L1-2}
&\langle \vec a + \nu \vec e, \vec Y|[L_2, V_\kappa(1)] |\vec b + (\xi + \nu+\mu) \vec e
,\vec W\rangle
+\frac{N \xi}{2}\langle \vec a + \nu \vec e, \vec Y|[J_2, V_\kappa(1)] |\vec b + (\xi + \nu+\mu) \vec e
,\vec W\rangle
 \\
&=\bigg\{\D \left(-\frac{\vec a + \nu \vec e}{\sqrt \b}-Q\vec \rho+Q\frac{N+1}{2}\vec e \right)+|\vec Y|
-\D\left(-\frac{\vec b + (\nu+\mu) \vec e}{\sqrt \b}-Q\vec \rho+Q\frac{N+1}{2}\vec e\right)-|\vec W|  \\
&
+\frac{(NQ-\kappa)^2}{N}
+{\kappa(\kappa-Q(N-1))}-\frac{\kappa^2}{N} \bigg\} Z(\vec a, \vec Y; \vec b, \vec W;\mu)  \\
&+{\sqrt \b}Q\langle \vec a + \nu \vec e, \vec Y|V_\kappa(1)J_1 |\vec b + (\xi + \nu+\mu) \vec e
,\vec W\rangle 
+{\sqrt \b}Q\langle \vec a + \nu \vec e, \vec Y|V_\kappa(1)J_2 |\vec b + (\xi + \nu+\mu) \vec e
,\vec W\rangle  \\
&+\sqrt {\beta}^{-1}\frac{N \xi}{2}U_{-2,0}Z(\vec a, \vec Y; \vec b, \vec W;\mu)\\
&=\sqrt{\beta}^{-1} U_{-2,1} Z(\vec a, \vec Y; \vec b, \vec W;\mu) 
+\sqrt {\beta}^{-1}\frac{N \xi}{2}U_{-2,0}Z(\vec a, \vec Y; \vec b, \vec W;\mu) \\
&+{\sqrt \b}Q\langle \vec a + \nu \vec e, \vec Y|V_\kappa(1)J_1 |\vec b + (\xi + \nu+\mu) \vec e,\vec W\rangle+{\sqrt \b}Q\langle \vec a + \nu \vec e, \vec Y|V_\kappa(1)J_2 |\vec b + (\xi + \nu+\mu) \vec e,\vec W\rangle
\end{split}
\end{equation}

Compare the above two equations, the Ward identity for $L_2$
is obtained since it is identified with the recursion formula 
$\delta_{-2,1} Z_{\vec Y,\vec W} -U_{-2,1} Z_{\vec Y,\vec W}=0$.
$L_{-2}$ totally follows the same discussion.


\end{document}